\documentstyle[12pt,psfig,epsf,epsfig,graphics,here,amssymb]{article}

\newcommand{\TEV}{\mbox{TeV}}
\newcommand{\MKM}{\mbox{$\mu$m}}

\def\be{\begin{equation}}
\def\ee{\end{equation}}
                         \def\bearr{\begin{eqnarray}}
                         \def\eearr{\end{eqnarray}}
\def\benum{\begin{enumerate}}
\def\eenum{\end{enumerate}}
\def\bitem{\begin{itemize}}
\def\eitem{\end{itemize}}

\def\ga{\gamma}
\def\gagah{\ga\ga \to h}
\def\gagaH{\ga\ga \to H}
\def\hbb{h \to b \bar b}
\def\Hbb{H \to b \bar b}

\def\mh{m_H}
\def\MA{M_A}
\def\tb{\tan\beta}
\def\mhmax{$m_h^{\rm max}$}
\def\msusy{M_{SUSY}}
\newcommand{\lsim}
{\;\raisebox{-.3em}{$\stackrel{\displaystyle <}{\sim}$}\;}
\newcommand{\gsim}
{\;\raisebox{-.3em}{$\stackrel{\displaystyle >}{\sim}$}\;}

\def\beq{\begin{equation}}
\def\eeq{\end{equation}}
\def\bea{\begin{eqnarray}}
\def\beaa{\begin{eqnarray*}}
\def\eea{\end{eqnarray}}
\def\eeaa{\end{eqnarray*}}
\def\bq{\begin{quote}}
\def\eq{\end{quote}}
\def\gappeq{\mathrel{\rlap {\raise.5ex\hbox{$>$}}
{\lower.5ex\hbox{$\sim$}}}}

\def\lappeq{\mathrel{\rlap{\raise.5ex\hbox{$<$}}
{\lower.5ex\hbox{$\sim$}}}}

\begin{document}

\begin{center}
EUROPEAN ORGANIZATION FOR NUCLEAR RESEARCH
\end{center}

\begin{flushright}
BNL-HET-01/32\\
CERN/PS Note-2001-062 (AE) \\
CERN-SL/2001-055 (AP) \\
CERN-TH/2001-235 \\
CLIC-Note 500 \\
NUHEP-EXP/01-050\\
UCRL-JC-145692     \\
\end{flushright}

\begin{center}
{\Large \bf 
Higgs Physics with a $\gamma \gamma$ Collider Based on CLIC~1}\\
\vspace*{1.0cm}
{ \bf D.~Asner$^1$, 
H.~Burkhardt$^2$,  
A.~De~Roeck$^2$,  
J.~Ellis$^2$, J.~Gronberg$^1$,  S.~Heinemeyer$^3$, 
M.~Schmitt$^4$, D.~Schulte$^2$, M.~Velasco$^4$} and  
{\bf F.~Zimmermann$^2$}\\
\vspace{1.0cm}
{\small 
$^1$ Lawrence Livermore  National Laboratory,  Livermore, California 
94550, USA\\
$^2$ CERN, CH-1211 Geneva 23, Switzerland  \\
$^3$ Brookhaven National Laboratory, Upton, New York, USA \\
$^4$ Northwestern University, Evanston, Illinois 60201, USA  \\
}
\vspace{1.0cm}
{\bf Abstract} \\
\end{center}

We present the machine parameters and physics capabilities of the CLIC
Higgs Experiment (CLICHE), a low-energy $\gamma\gamma$ collider based on
CLIC~1, the demonstration project for the higher-energy two-beam
accelerator CLIC. CLICHE is conceived as a factory capable of producing
around 20,000 light Higgs bosons per year. We discuss the requirements for
the CLIC~1 beams and a laser backscattering system capable of producing
a $\gamma\gamma$ total~(peak) luminosity of $2.0~(0.36) \times 
10^{34}$~cm$^{-2}$s$^{-1}$
with $E_{CM}(\gamma \gamma) \sim 115$~GeV. We show how CLICHE could be
used to measure accurately the mass, $\bar b b$, $WW$ and $\gamma \gamma$
decays of a light Higgs boson. We illustrate how these measurements may
distinguish between the Standard Model Higgs boson and those in
supersymmetric and more general two-Higgs-doublet models, complementing
the measurements to be made with other accelerators. We also comment on
other prospects in $\gamma\gamma$ and $e^- \gamma$ physics with CLICHE.


\newpage
\section{Introduction}

CLIC~\cite{c:clic} is a project for a multi-TeV linear $e^+ e^-$ collider
using an innovative two-beam acceleration technique to achieve a high
accelerating gradient, and CLIC~1~\cite{c:clic1} 
is an essential step proposed at CERN in
the R\&D programme towards CLIC. CLIC~1 would be based on one module of
the eventual full CLIC accelerator, capable of delivering an intense,
low-emittance $e^-$ beam with an energy $\sim 70$~GeV. CLIC~1 is required
to provide proof-of-principle for the two-beam acceleration mechanism of
CLIC on a large scale, and would represent a major investment in
engineering and other resources. It is natural to seek to maximize the
physics return on this investment.

Several possible uses of an intense $e^-$ beam are readily apparent.
One could in principle use it for {\it fixed-target experiments},
for example. One could envision {\it colliding CLIC~1
with the LHC proton beam}, but the orientation and depth of CLIC~1 that
would be required are not compatible with the possibility of extending
CLIC~1 subsequently to become the full CLIC machine. If one had two CLIC~1
machines, a positron source and damping rings, one could make $e^+ e^-$
collisions. There has recently been a resurgence of physics interest in a
new round of high-statistics experiments at the $Z^0$ peak, 
{\it GigaZ}~\cite{TESLA_TDR,Orange,gigaz},
aimed at higher-precision electroweak measurements. The centre-of-mass
energy range $\sim m_Z$ would certainly be within range of a pair of
CLIC~1 machines, but effectiveness of the {\it GigaZ} programme would require
the positrons to be polarized, as well as the electrons, which is a
technical challenge. There is also interest in returning to {\it the $e^+
e^- \rightarrow W^+ W^-$ threshold}, in order to measure $m_W$ more
precisely. The $ W^+ W^-$ threshold would be within reach of a modest
upgrade of the nominal CLIC~1 energy.

Alternatively, with two CLIC~1 machines, or even just one feeding two arcs
{\`a la} SLC, and a laser backscattering laser facility to 
produce high-energy photons by the Compton process   
$e^- \gamma_{laser} \rightarrow   e^- \gamma$.
One could make {\it $e^- \gamma$ and/or $\gamma \gamma$
collisions} at centre-of-mass energies up to $\sim 0.9$ or $\sim 0.8
\times E_{CM}(e^-e^-)$, respectively. A polarized positron source would
not be needed for such experiments. The principles of photon colliders
\cite{Telnov_et_al} and the physics interest of $\gamma \gamma$
collisions, in particular, have been documented
extensively~\cite{nlc_report,teslatdr}.

Precision electroweak fits suggest~\cite{lepEWWG} that the most likely 
mass for the Standard Model Higgs
boson is just above the limit of 114~GeV provided by direct
searches at LEP~\cite{LEPHWG}. Moreover, the mass range $m_H \lappeq 
130$~GeV~\cite{mh130} is
suggested independently by supersymmetry.  We also recall that the last
days of LEP even provided a direct hint for a Standard Model-like Higgs
boson with mass $\sim 115$~GeV~\cite{LEPHWG}, and that LEP could not 
exclude a
supersymmetric Higgs boson as light as $\sim 90$~GeV. The possibility of
such a light Higgs boson may be confirmed or refuted by the Tevatron
collider within a few years, and the LHC would certainly measure the value
of $m_H$. With the injection energy of 9~GeV, the nominal energy
of CLIC~1 is 77~GeV. Therefore, the effective $E_{CM}(\gamma
\gamma)$ could be tuned to a value up to $\sim 0.8 \times 
154$~GeV $\sim 120$~GeV.

Thus there exists an opportunity for turning CLIC~1 into a Higgs factory,
a proposal we term the CLIC Higgs Experiment, or CLICHE. As discussed in
~\cite{c:fd}, the option of colliding the beams from two CLIC~1 machines 
looks
promising, and the geometric $e^-e^-$ luminosity may attain around  $4.8
\times
10^{34}$~cm$^{-2}$s$^{-1}$. Combined with a suitable laser backscattering
system, such a luminosity would enable accurate measurements of
the properties of the Higgs boson. For example, the laser system described
below could yield a total $\gamma \gamma$ luminosity of up to
200~fb$^{-1}$ per `Snowmass year'  of $10^7$s, which could produce around 
22,000 light 
Standard Model (SM) Higgs
particles.

In subsequent sections of this paper, we first assemble some initial
considerations of accelerator aspects of CLICHE, including the
requirements for the CLIC~1 beam parameters and the laser system if one is 
to
attain a luminosity sufficient to study Higgs physics. Then we review
briefly some of the most interesting physics measurements possible with
CLICHE, which could include accurate measurements of $m_H$, $\Gamma(H
\rightarrow \gamma \gamma)\times {\cal B}r(H\rightarrow {\bar b}b)$,
$\Gamma(H \rightarrow \gamma \gamma)\times {\cal B}r(H\rightarrow WW)$,
$\Gamma(H \rightarrow \gamma \gamma)\times {\cal B}r(H\rightarrow
\gamma\gamma)$, and the CP properties of the $ H \rightarrow \gamma
\gamma$ coupling. We study the capability of CLICHE to distinguish a
Standard Model Higgs from the lightest Higgs boson in the minimal
supersymmetric extension of the Standard Model (MSSM), and also discuss a
more general two-Higgs-doublet model.

As we also mention, there are other physics processes that might be
interesting at CLICHE, {\it e.g.}, QCD reactions in $\gamma \gamma$
collisions and the reaction $e^- \gamma \rightarrow \nu W^-$ in $e^-
\gamma$ collisions, which would provide an opportunity to measure $m_W$
and $\Gamma_W$. We also advertise the physics opportunities for Higgs
physics offered by higher-energy $\gamma \gamma$ colliders, for which
CLICHE might serve as an engineering prototype. For example, the study of
higher-mass Higgs bosons may best be done at high-energy
$\gamma\gamma$ colliders, due to the fact that they are produced in the
$s$ channel (so that all of the phase space is available for producing the
Higgs mass)~\cite{muhl,david_jack}, and the fact that the photon beams can be produced in a state
of definite CP.

\section{Accelerator Considerations}

This section describes two important components of 
CLICHE, namely the CLIC~1 electron accelerator and the laser 
backscattering system. 

At CERN, a high energy, high luminosity electron-positron 
linear collider (CLIC for Compact LInear Collider) is being 
studied as a possible post-LHC facility~\cite{c:clic}.
It is based on a novel two-beam scheme and uses high-frequency,
high-gradient normal-conducting structures to accelerate the beam.
The power to accelerate this main beam is obtained by decelerating
drive beams in a dedicated beam line passing 
parallel to the main linac. CLIC requires two
drive-beam complexes, each of which generate the 22 drive-beam pulses
necessary to power 22 drive-beam decelerators for one of the two 
main linacs, for e$^{+}$e$^{-}$ collisions at 3 TeV centre-of-mass
energy. The two-beam acceleration has been demonstrated 
successfully in two test facilities (CTF1 and CTF2). A test of 
the drive-beam generation will take place
at a new test facility (CTF3)~\cite{CTF3}, presently 
under construction. Its operation will enables one to judge whether the 
technology works, and it is expected that
a conceptual design report for CLIC could be completed by the end of 2006.
CTF3 should be followed by CLIC~1, which is conceived to provide 
a full scale test of beam dynamics and power handling.

\begin{figure}
\begin{center}
\epsfbox{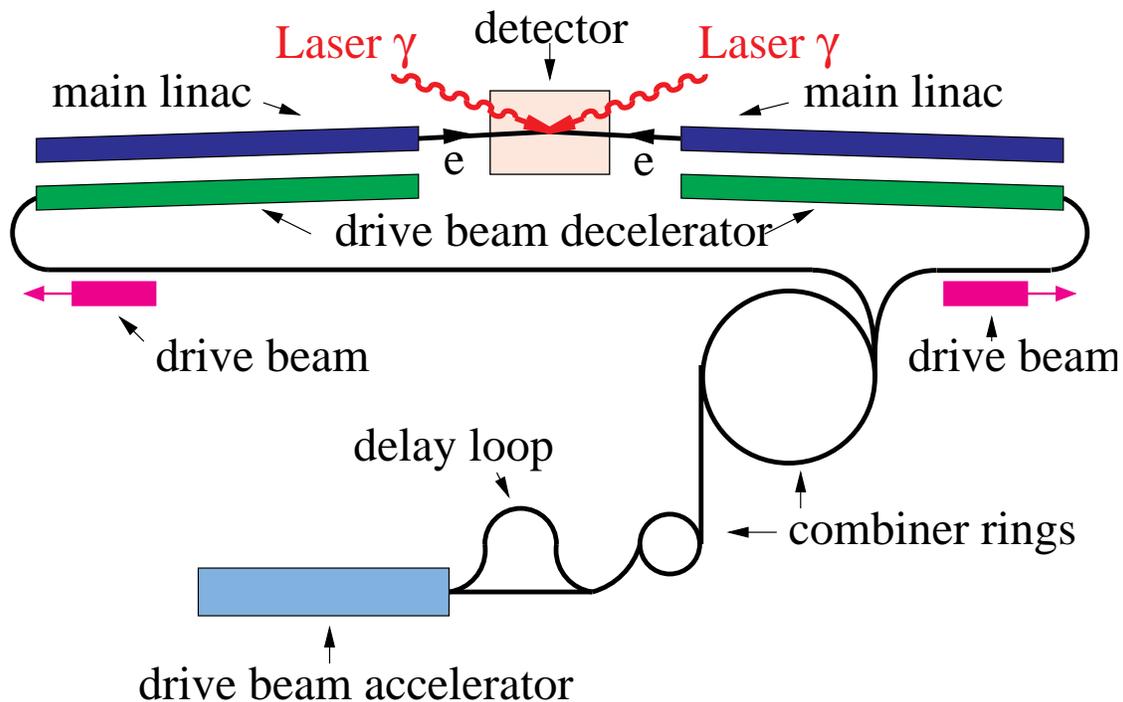}
\end{center}
\caption{\it Schematic sketch of a layout for a $\gamma\gamma$
collider based on CLIC~1.}
\label{f:scheme}
\end{figure}

\subsection{Parameters of CLIC~1}

CLIC~1 is proposed to consist of one drive-beam generation complex and one
drive-beam decelerator, with the corresponding
length of main linac sufficient to accelerate a main beam by about
68~GeV.

Recently, an exploratory study has been carried out to determine how this
facility could be turned into a collider with a high geometric $\rm
e^-e^-$ luminosity, which could be used as the basis for a $\gamma\gamma$
collider~\cite{c:fd}. This would require the addition of a small-emittance
main-beam source and a final-focus system able to achieve small spot sizes
at the interaction point (IP). In addition, it would be necessary to
achieve a high beam current in the main linac. The main linac of CLIC~1
could in principle be turned into a collider by adding arcs as in
SLC~\cite{c:slc}. However, it seems simpler to add a second CLIC~1 linac
pointing at the first one, and only this option is considered here. Some
preliminary parameters of a potential $\rm e^-e^-$ collider are given in
Table~\ref{t:gamma}. A more detailed investigation is required in
order to establish the feasibility of the approach and to find possible
improvements.

In the proposed CLIC~1 collider scheme, the CLIC drive-beam complex can
power both linacs. Alternate pulses are sent into the first and second
linac, as seen in Fig.~\ref{f:scheme}. The two pulses accelerate the main
beams at the same time if the drive-beam complex is properly placed.
The third pulse can be used to power again the
first linac and the fourth to power the second, and so on, using all
drive-beam
pulses. This effectively increases the repetition frequency from $100\,\rm
Hz$ to $1.1\,\rm kHz$. However, it remains to be investigated if the
structures can tolerate the heat load associated with this mode of
operation. A repetition frequency of $200\,\rm Hz$ is foreseen for 
CLIC at a centre-of-mass energy of 500~GeV. The effective 
repetition frequency that we assume
is therefore in the range of $200$--$1100\,\rm Hz$
depending on the heat load capabilities of the structures. This
leads to an uncertainty in the final luminosity by a factor of five.

In the preferred two-beam scheme, the main beams would be generated by a
conventional electron source, which could provide a polarization of about
80\%. The injector could likely use the slightly modified SPS as a damping
ring~\cite{c:sps,c:jowett}. The beams would be accelerated to 
9~GeV before 
injection into the main linac, increasing the maximum beam
energy to $\sim 77$~GeV. After the linac, a collimation system would
scrape off the beam tails, and a final-focus system would focus the beams
down to about $2.9\,\rm nm$ vertically and $154\,\rm nm$
horizontally.

The geometric $\rm e^-e^-$ luminosity that may be achievable is 
${\cal L}\simeq (4 \times10^9 \times 4 \times10^9/154  \times10^{-7}/2.9 
\times10^{-7})(100\times 11\times 154/4/3.1415)=4.8 \times10^{34}
\rm cm^{-2}s^{-1}$ and 
${\cal L}\simeq 0.9 \times10^{34}\,\rm cm^{-2}s^{-1}$ for an effective
repetition rate of  $1.1\,\rm kHz$ and $200\,\rm Hz$, respectively. It may 
be possible to increased the luminosity by using a larger bunch charge and 
by
decreasing the vertical emittance, and further study is needed to find the 
limits of these parameters.

\begin{table}[htbp]
\caption{\it Example parameters for a CLIC~1 collider.}
\label{t:gamma}
\begin{center}
\begin{tabular}{lccc}
\hline
variable & symbol &\multicolumn{2}{c}{value} \\
\hline
total power consumption for RF&
$P$ &\multicolumn{2}{c}{150 MW} \\
beam energy & $E$ &\multicolumn{2}{c}{75 GeV}\\
beam polarization & $P_e$ &\multicolumn{2}{c}{0.80}\\
bunch population & $N$ &\multicolumn{2}{c}{$4\times 10^{9}$}\\
number of bunches per train & $n_{b}$ &\multicolumn{2}{c}{154}\\
number of trains per rf pulse & $n_{t}$ & \multicolumn{2}{c}{11} \\
repetition rate & $f_{\rm rep}$ &\multicolumn{2}{c}{100 Hz}\\
rms bunch length & $\sigma_{z}$ &\multicolumn{2}{c}{30 $\mu$m}\\
crossing angle & $\theta_{c}$ &\multicolumn{2}{c}{$\ge 20$ mrad} \\
normalised horizontal emittance & 
$\epsilon_{x}$ &\multicolumn{2}{c}{1.4\,$\mu$m}\\
normalised vertical emittance & 
$\epsilon_{y}$ &\multicolumn{2}{c}{0.05\,$\mu$m}\\
nominal horizontal beta function at the IP &
$\beta_{x}^{\ast}$ &\multicolumn{2}{c}{2\,mm} \\
nominal vertical beta function at the IP &
$\beta_{y}^{\ast}$ &\multicolumn{2}{c}{20\,$\mu$m} \\
e$^{-}$e$^{-}$ geometric luminosity & ${\cal L}$ &
\multicolumn{2}{c}{$0.9$--$4.8 \times 10^{34}$~cm$^{-2}$s$^{-1}$} \\
\hline
\end{tabular}
\end{center}
\end{table}

\subsection{Laser Backscattering System}

The photon beams required by CLICHE would be produced via the Compton
backscattering of laser light off the high-energy electron beam from
CLIC~1. The bunch-to-bunch distance and pulse length of CLICHE are of the
same
order of magnitude as those for NLC.
The laser requirements for CLIC and NLC are 
therefore comparable, since they both use all the Mercury
technologies~\footnote{This  laser technology is 
capable of delivering the 10 kW of average power in short pulses
of 1 TW peak power. The total energy of a pulse is 1~J.} - pump diodes, Yb-SFAP crystals, cooling, chirp pulse, 
etc.~\cite{nlc_report},
the only 
differences being at the front end. On the other hand, the TESLA 
option~\cite{teslatdr} 
requires a high-speed lower-power optical switch (Pockels cell) which is 
under development at LLNL.  

In the laser-beam collision at the conversion point,
the maximum energy of the scattered photons is:
\begin{equation}
\omega_m=\frac{x}{x+1}E_0; \;\;\;\;
x \approx \frac{4E_0\omega_0}{m^2c^4}
 \simeq 15.3\left[\frac{E_0}{\TEV}\right]
\left[\frac{\omega_0}{eV}\right],
\end{equation}
where $E_0$ is the electron beam energy and $\omega_0$ the energy of the
laser photon. In connection with NLC studies~\cite{nlc_report}, the case
$E_0 =250$~GeV,
$\omega_0 =1.17$~eV, i.e.,
$\lambda=1.0$ \MKM\ , has been considered. This 
would
correspond to $x=4.5$ and $\omega_m = 0.82E_0$. In the case of CLICHE,
the centre-of-mass energy of the accelerator would be
$E_{CM}({\rm e}^-{\rm e}^-)\simeq~150$~GeV. 
In order for $E_{CM}(\gamma \gamma)$ to be close to the mass of a
115~GeV Higgs boson, the energy of the laser photon is chosen
to be 3.53~eV instead of 1.17~eV, resulting in a maximum photon energy of 
roughly 60~GeV. 

We need to add frequency multipliers to reduce the wavelength. The 
increase in the
laser frequency is achieved by adding a tripler to the laser system.  In
this case it can be assumed that the 1.054~$\mu$m laser can be turned into
one with 0.351~$\mu$m. As a consequence, the power provided by the photons 
is reduced
by a factor 1/3. However the area of the focal spot is also reduced by
1/3, so the overall effect cancels out in the power in the peak. A 70\% 
tripling efficiency is expected. Parameters of the
electron and laser beams are shown in Table~\ref{t:gamma} and ~\ref{t:gamma2}.
We find that CLICHE
requires between $154 \times 2 \times 100 = 30800$ and
$154 \times 11 \times 100 = 169400$ pulses/second. 
As a consequence, 
the use of more elaborate multi-pass optics than the
two-pass system designed for NLC~\cite{nlc_report} would be important
for reducing the required number of laser pulses.

\begin{table}[htbp]
\caption{\it Example of laser parameters for a $\gamma\gamma$
collider based on CLIC~1  
for ${{\cal L}_{ee}=4.8 \times 10^{34}\rm cm^{-2}s^{-1}}$.}
\label{t:gamma2}
\begin{center}
\begin{tabular}{lccc}
\hline
variable & symbol &\multicolumn{2}{c}{value} \\
\hline
{Laser beam parameters} & & & \\
\hline
  Wavelength        & $\lambda_L$ & \multicolumn{2}{c}{0.351  $\mu$m} \\
  Photon energy & $\hbar\omega_L$ & \multicolumn{2}{c}{3.53  eV = 5.65$\times 10^{-19}$ J} \\
  Number of laser pulses per second & $N_L$ &
\multicolumn{2}{c}{169400\,s$^{-1}$} \\
  Laser peak power    & $W_L$ &
\multicolumn{2}{c}{2.96$\times 10^{22}$ W/m$^2$} \\
  Laser peak photon density    &       &
\multicolumn{2}{c}{5.24$\times 10^{40}$ photons/m$^2$/s} \\
\hline
{Photon beam} & &  \\
\hline
Number of photons per electron bunch & $N_{\gamma}$ &
\multicolumn{2}{c}{$9.6\times10^{9}$} \\
$\gamma\gamma$ luminosity & ${\cal L}_{\gamma\gamma}$ &
                  \multicolumn{2}{c}{$2.0\times10^{34}$
cm$^{-2}$s$^{-1}$} \\
$\gamma\gamma$ luminosity for $E_{\gamma\gamma}\ge 0.6 E_{CM}$ & ${\cal
L}_{\gamma\gamma}^{peak}$ &
                  \multicolumn{2}{c}{$3.6\times10^{33}$ cm$^{-2}$s$^{-1}$}
\\
\hline
\end{tabular}
\end{center}
\end{table}

The optimum polarization combination that we anticipate 
for the electron beam $P_e$ and the laser beam $P_L$
is $P_e P_L=-0.8$. In this case the generated photon spectrum
peaks at its maximum energy, $\omega_m$. In addition,
the high-energy photon beam is almost
completely polarized around the peak energy.
In order to enhance the Higgs production and to suppress the
background events, from $\gamma \gamma \rightarrow {\bar b} b + n g$ in
particular, the polarizations of the colliding photon beams should be
arranged so that $J_z=0$ collisions dominate. With the CLIC~1 beam energy,
the Higgs particle of 115~GeV is produced almost at rest, while the
low-energy background events are strongly boosted. Hence the backgrounds
will in general have topologies that are different from the desired signal
events, as discussed later.

\begin{figure}[t]
\begin{center}
\mbox{\epsfig{file=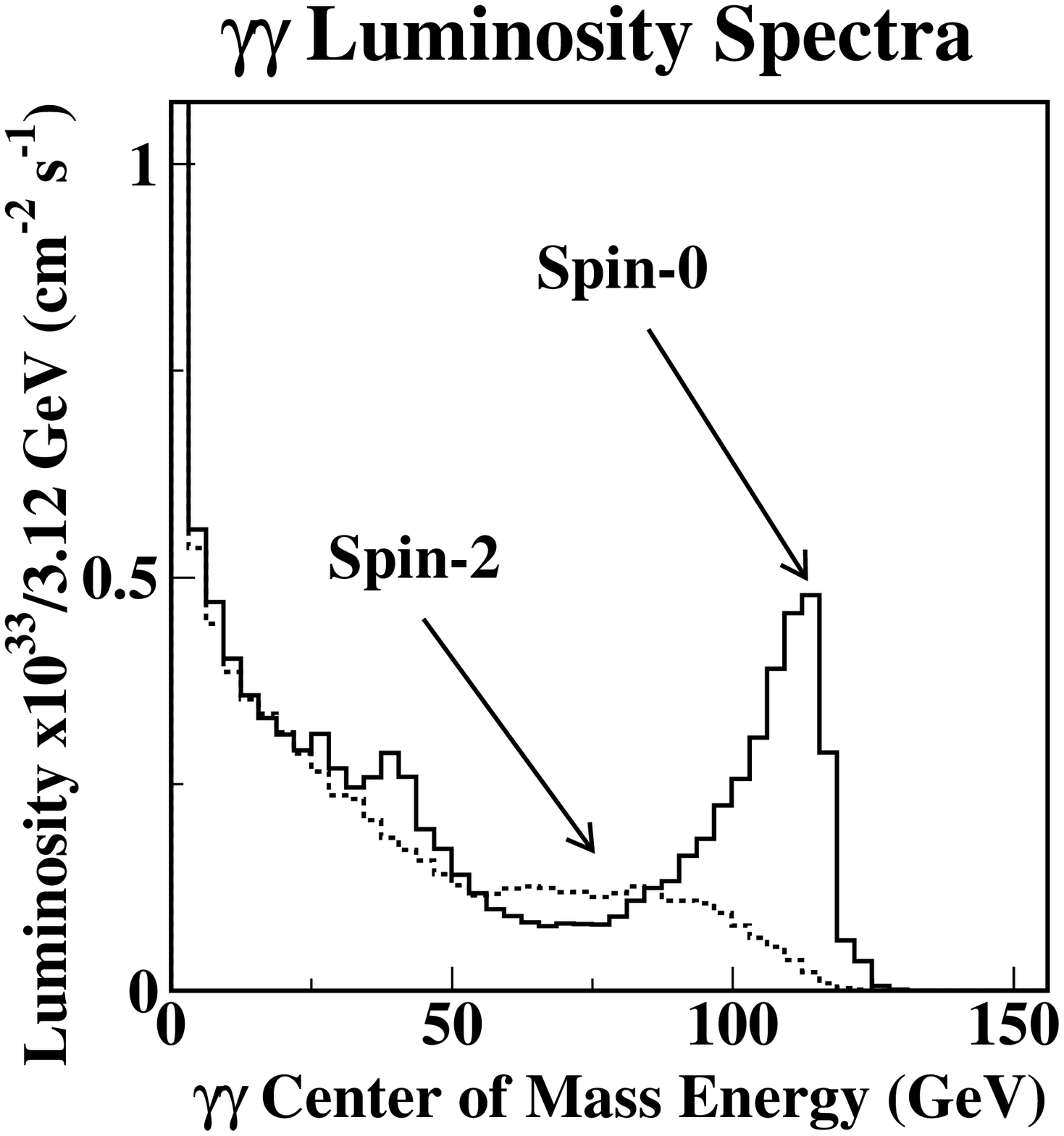,height=8cm}}
\mbox{\epsfig{file=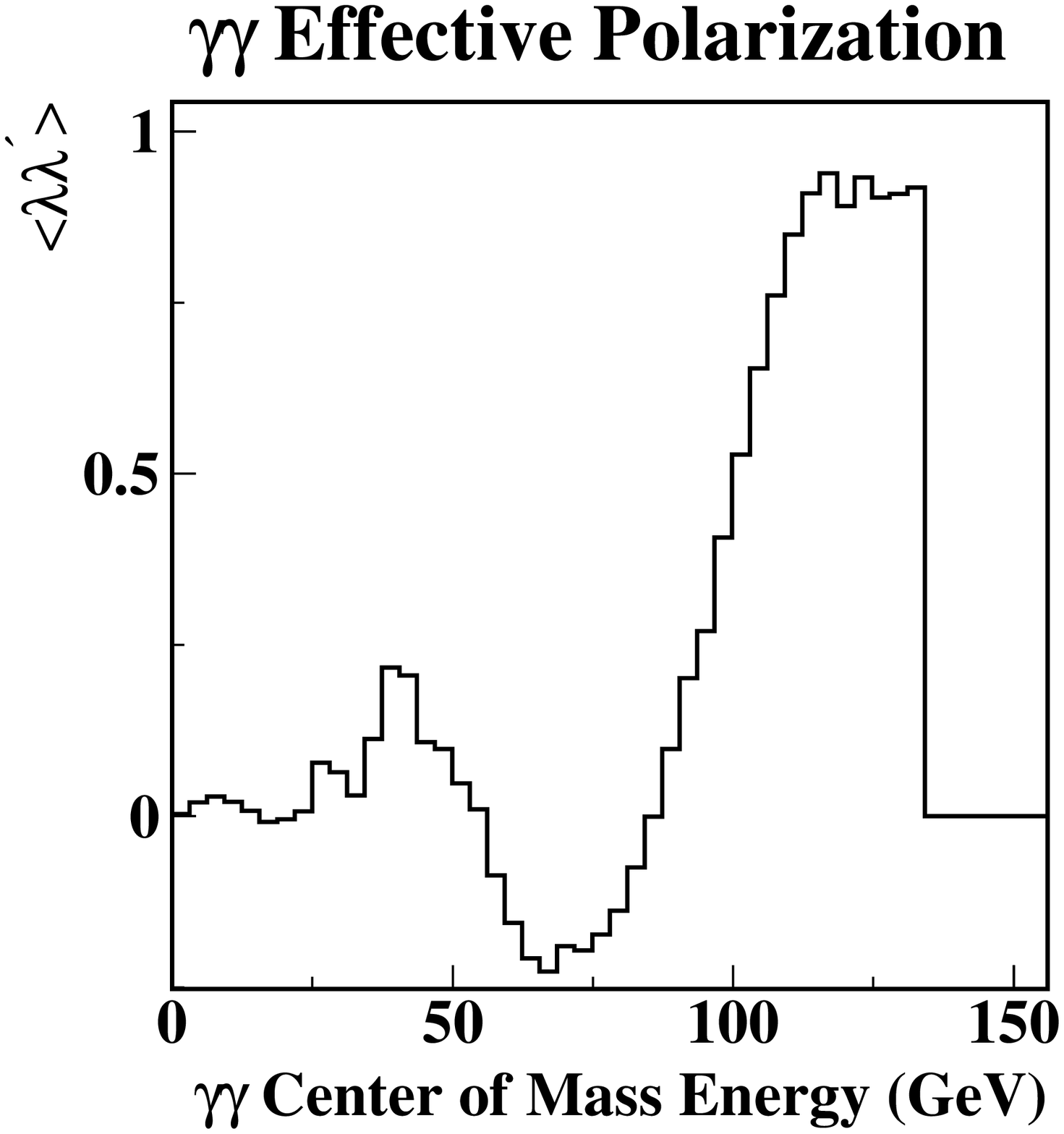,height=8cm}}
\end{center}
\caption[.]{\label{fig:spectra}\it
Luminosity spectra and beam polarization as functions of
$E_{CM}$($\gamma\gamma$) for the CLIC~1 parameters for 75~GeV electrons
obtained with {\tt DIMAD}~\cite{dimad} and {\tt CAIN}~\cite{cain2} for 
 ${{\cal L}_{ee}=4.8 \times10^{34} \rm cm^{-2}s^{-1}}$.

}
\end{figure}

The expected total energy spectra and  polarization are 
shown in Fig.~\ref{fig:spectra}.  
The luminosity calculation assumes improvements over what was
originally proposed in~\cite{c:fd}:
\begin{itemize}

\item  use of the most recent NLC Final-Focus System~(FFS) for operation
of the center-of-mass of 0.5-1 TeV~\cite{c:nlc_ffs}, 

\item  and increasing the strength of the bending magnets by a factor of 
4.35, with corresponding decreases in the sextupole, octupole and 
decapole strengths by 4.35, 4.35$^2$ and 4.35$^3$.
This allows the beta functions to be reduced from the NLC values:
$\beta_{x}$=2\,mm (previously 8\,mm) and $\beta_{y}$=20\,$\mu$m 
(previously 100\,$\mu$m).
\end{itemize}

Tracking simulations of the FFS that take into account the
synchrotron radiation in the dipoles and quadrupoles and the initial beam
energy distribution from the linac as simulated with {\tt DIMAD}~\cite{dimad} 
and {\tt PLACET}~\cite{c:placet} predict
a geometric luminosity of roughly 4.8$\times 10^{34}$~cm$^{-2}$s$^{-1}$ 
for 11 trains per rf pulse.
The energy spread was assumed to be $\Delta E/E\sim 0.23\%$. 
The horizontal and vertical normalized emittances were taken to be 
1.4$\times 10^{-6}$~m and 5$\times 10^{-8}$~m, respectively, and the RMS
at the IP is 138.1~nm~(2.6~nm) in the horizontal~(vertical) plane. However,
{\tt DIMAD} predicts that the effective beam spot is larger,
($154\times 2.9$)~nm$^2$.  This is due to effects such as synchrotron radiation 
and the combination of chromaticity of the FFS and the electron
 beam energy  spread. 
As a consequence, the luminosity is simulated to be 15\% lower than the `ideal' 
scenario.  These effects  are  taken into account in the luminosity
results shown in Fig.~\ref{fig:spectra}.  Using the laser parameters described 
above, the $\gamma\gamma$ luminosity with a centre-of-mass energy above
$0.6\times 150$~GeV
is about $3.6 \times 10^{33}$cm$^{-2}$s$^{-1}$, and the polarization
$<\lambda\lambda^\prime >$ at the peak is 0.94.

We have chosen the distance between the conversion point and the interaction
point to be 1mm. Detailed studies indicate that the optimal value is 
1.4 mm, which would yield a somewhat higher luminosity. On the other hand,
there would be a reduction 
in luminosity if a more conservative beam spot size were assumed, so we 
consider the figures presented here quite representative.

\begin{figure}[ht]
\begin{center}
\mbox{\epsfig{file=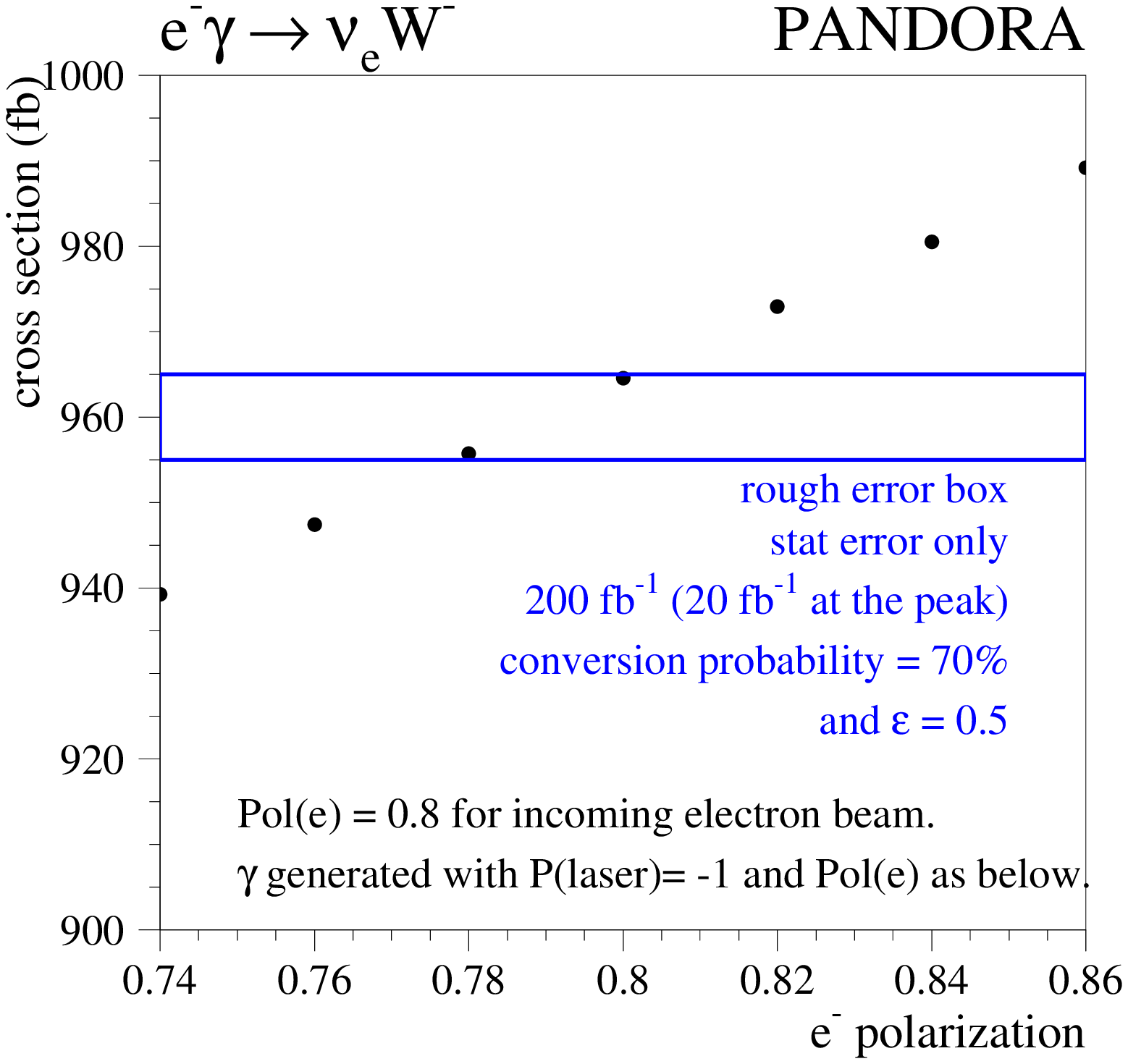,height=10cm}}
\end{center}
\caption[.]{\label{fig:polarizetion}\it
The variation in the cross section for the $e\gamma \rightarrow W \nu$
process as a function of polarization. This analysis includes the full
photon spectrum in the cross section calculation. The attainable
statistical error in the cross-section measurement is also indicated.
}
\end{figure}

The laser-beam collision at the conversion point and the beam-beam
collision at the interaction point have been simulated using {\tt 
CAIN}~\cite{cain2} and using {\tt GUINEA-PIG}~\cite{c:guinea}.
The results of the two programs agree quite well.  The 
luminosity spectra and the effective beam polarization as functions of    
$E_{CM}$($\gamma\gamma$) obtained using {\tt CAIN} are shown in 
Fig.~\ref{fig:spectra}.
Both of these need to be monitored and
controlled accurately. At the E$_{CM}$ under consideration,
the spectra and luminosity can be measured using the reaction
$\gamma\gamma\rightarrow e^+e^-$~\cite{spectra_undestanding}. The reaction 
$\gamma\gamma\rightarrow e^+e^- \gamma$ may also be useful for this
purpose, 
but this requires further study. The polarization cannot be measured using 
the reaction $\gamma\gamma\rightarrow \mu^+\mu^-\mu^+\mu^-$ as originally
suggested~\cite{spectra_undestanding}. A promising way of measuring the
photon polarization uses the reactions $e\gamma \rightarrow e\gamma$ and
$e\gamma \rightarrow W \nu$. A detailed study of the second process in
which the full photon energy spectrum is taken into account is in   
progress~\cite{ongoing}. Fig.~\ref{fig:polarizetion} displays the variation
of the cross section with the photon polarization, together with the
attainable statistical error. The latter would correspond to a determination of
the photon polarization with a precision of $\pm 1$\%.

\section{Physics Opportunities}

Our primary thrust in this paper is to emphasize the Higgs physics accessible
with CLICHE. However, there are  other interesting physics opportunities, as we
discuss in the last two subsections.

\subsection{A Higgs Factory}

There have been many investigations of the physics possibilities of
$\gamma\gamma$ colliders, including discussions of $\gamma\gamma$~options
at JLC~\cite{jlc},  NLC~\cite{nlc_report} and  TESLA~\cite{teslatdr}.
Here we describe results from exploratory studies of a $\gamma\gamma$
collider optimized for a light  Higgs factory scenario at CLIC~1.
As discussed in some detail below, several important measurements
of Higgs properties can be made at a Higgs factory.  In fact, for many 
of the analyses, running close to the Higgs threshold has important advantages.

\begin{figure}[h]
\begin{center}
\resizebox{\textwidth}{!}
{\epsfig{file=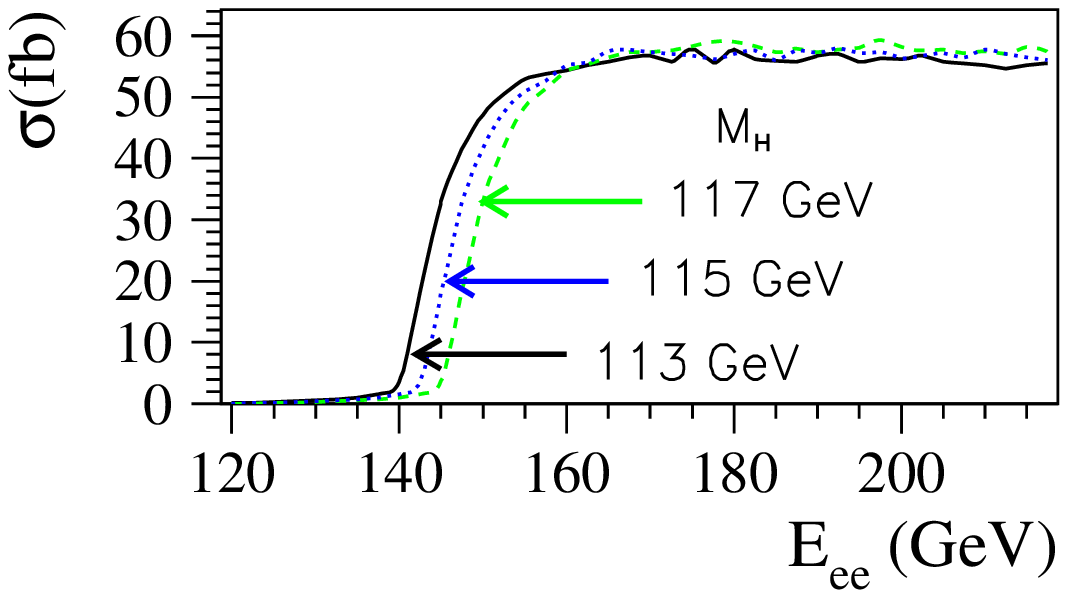,height=6.5cm}
\epsfig{file=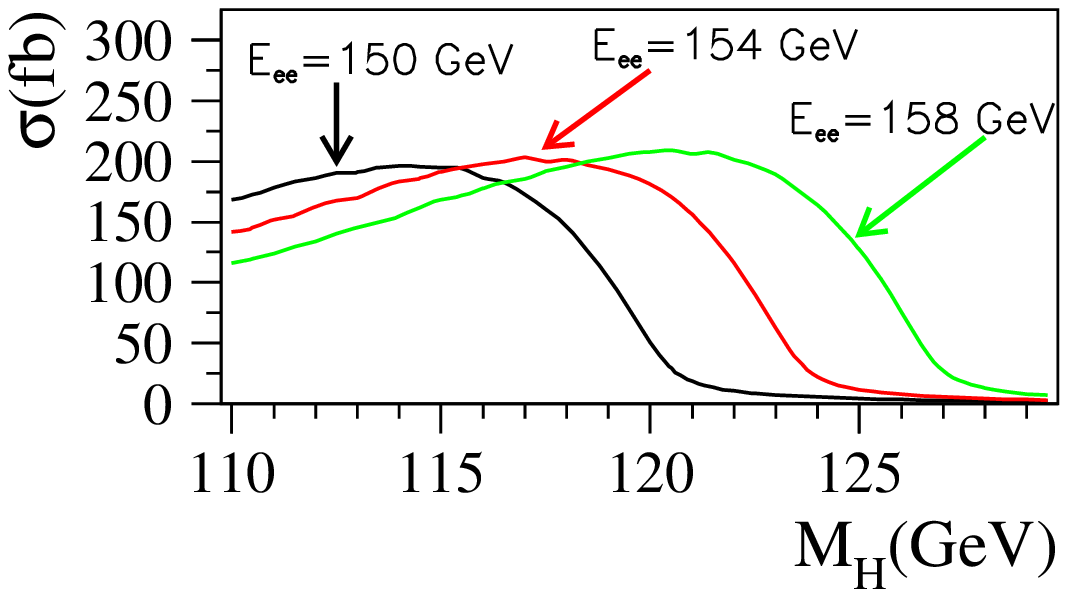,height=6.5cm}}
\end{center}
\caption[.]{\label{fig:excitation}\em
(a) The cross sections for $\gamma \gamma \rightarrow H$ for 
different values of $m_H$ as functions of $E_{CM}(e^- 
e^-)$ for unpolarized photons. 
(b) The cross section for $\gamma \gamma \rightarrow H$
as a function of $m_H$ for three different values of  $E_{CM}(e^- e^-)$.
Here the electrons are assumed to be 80\% polarized longitudinally,
and the lasers circularly polarized, so that the produced 
photons are highly circularly polarized at their maximum energy. 
}
\end{figure}

We begin with a discussion of the Higgs production cross section.
The excitation curve for a Higgs boson with mass  around $115$~GeV as a 
function of
$E_{CM}(e^- e^-)$ for unpolarized electrons is shown in
Fig.~\ref{fig:excitation}(a). We see that the cross section rises rapidly
for $E_{CM}(e^- e^-)$ between 140 and 160~GeV, providing a physics
opportunity for CLIC~1 with a beam energy of 77~GeV,
if indeed $m_H \sim 115$~GeV. Fig.~\ref{fig:excitation}(b) shows the cross
section as a function of Higgs mass for three choices of $E_{CM}(e^-
e^-)$. We note that the excitation curve increases by a factor of three if
the electron beams are 80\% longitudinally polarized. The
CLIC~1 energy could be somewhat lower if the lower limit of about 90~GeV 
on
the lightest MSSM
Higgs mass is saturated, whereas an energy upgrade would be required if
the MSSM upper bound of about 130~GeV were to be saturated.

\begin{figure}[htbp]
\begin{center}
\mbox{\epsfig{file=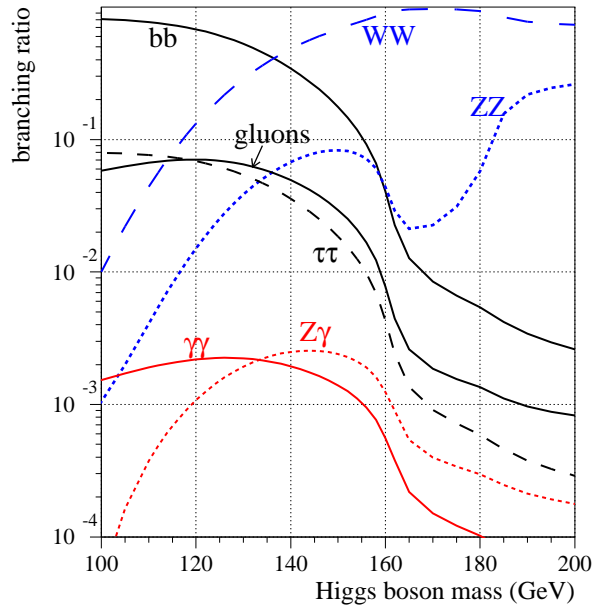,height=8cm}}
\end{center}
\caption[.]{\label{fig:decaymodes}
\em
The principal decay modes of the Higgs boson~\cite{hdecay} in the mass 
range favoured by precision electroweak experiments.
}
\end{figure}

The most important decay modes of a Standard Model Higgs boson in the
range between 100 and 200~GeV are shown in Fig.~\ref{fig:decaymodes}. We
see that the $H \rightarrow \gamma \gamma$ coupling used to produce the
Higgs boson at CLICHE yields a relatively minor decay mode, whereas the
dominant decay mode for a Standard Model Higgs boson weighing $\sim
115$~GeV is $ H \rightarrow {\bar b} b$. Indeed, the most promising
reaction at such a Higgs factory is $\gamma \gamma \rightarrow H
\rightarrow {\bar b} b$, as we discuss below. Other decay channels
accessible at CLICHE include $H \rightarrow W W$ and $H \rightarrow
\gamma\gamma$.  In the Standard Model, the branching ratios for ${\cal
B}r(H \to b\bar{b})$, ${\cal B}r(H \to WW)$ and ${\cal B}r(H \to
\gamma\gamma)$ for a Higgs mass of 115~GeV are:  73.7\%, 8.8\% and 
0.2\%,
respectively. One of the objectives of CLICHE would be to test these
predictions, and use measurements of them to distinguish between the
Standard Model and its possible extensions, such as the minimal
supersymmetric extension of the Standard Model (MSSM) or a more general
two-Higgs-doublet model (2HDM), as we discuss later.

\subsubsection{Higgs measurements with CLICHE}

In all the studies shown below, realistic beam spectra and luminosities
were used. The calculations were based on the {\tt CAIN } program~\cite{cain2}
which includes the non-linear effects that can cause  distortions in
the photon beam energy spectra, as well as
the latest description of the interaction region. The photon 
helicities were taken into account in the signal and background estimations. 
The events were generated using 
{\tt PANDORA-PYTHIA}~\cite{pythiajetset,pandora}, 
and {\tt LCDROOT FASTMC}~\cite{fastmc} was used for the detector simulations, 
including calorimeter energy  smearing.

We give only a brief overview of exploratory results 
pertaining to the Standard Model Higgs boson
with a mass in the range $110$--$125$~GeV.

\noindent\underline{Mass}\\
~\\
A special feature of the $\gamma\gamma$ collider is the sharp edge of
the $\gamma\gamma$ luminosity function, as depicted in Fig.~\ref{fig:spectra}.
The position of this edge can be controlled by changing the electron beam
energy.  As it sweeps across the threshold for Higgs production, the 
number of, e.g., $\bar{b}b$ events will increase dramatically.  This
phenomenon is already reflected in the sharpness of the excitation
curves of Fig.~\ref{fig:excitation}.  Since the position of this turn-on
depends on the Higgs mass, a threshold scan offers the possibility to
measure the Higgs mass kinematically, as developed in~\cite{ohgaki}.
%

\begin{figure}[htbp]
\begin{center}
\resizebox{\textwidth}{!}
{\epsfig{file= 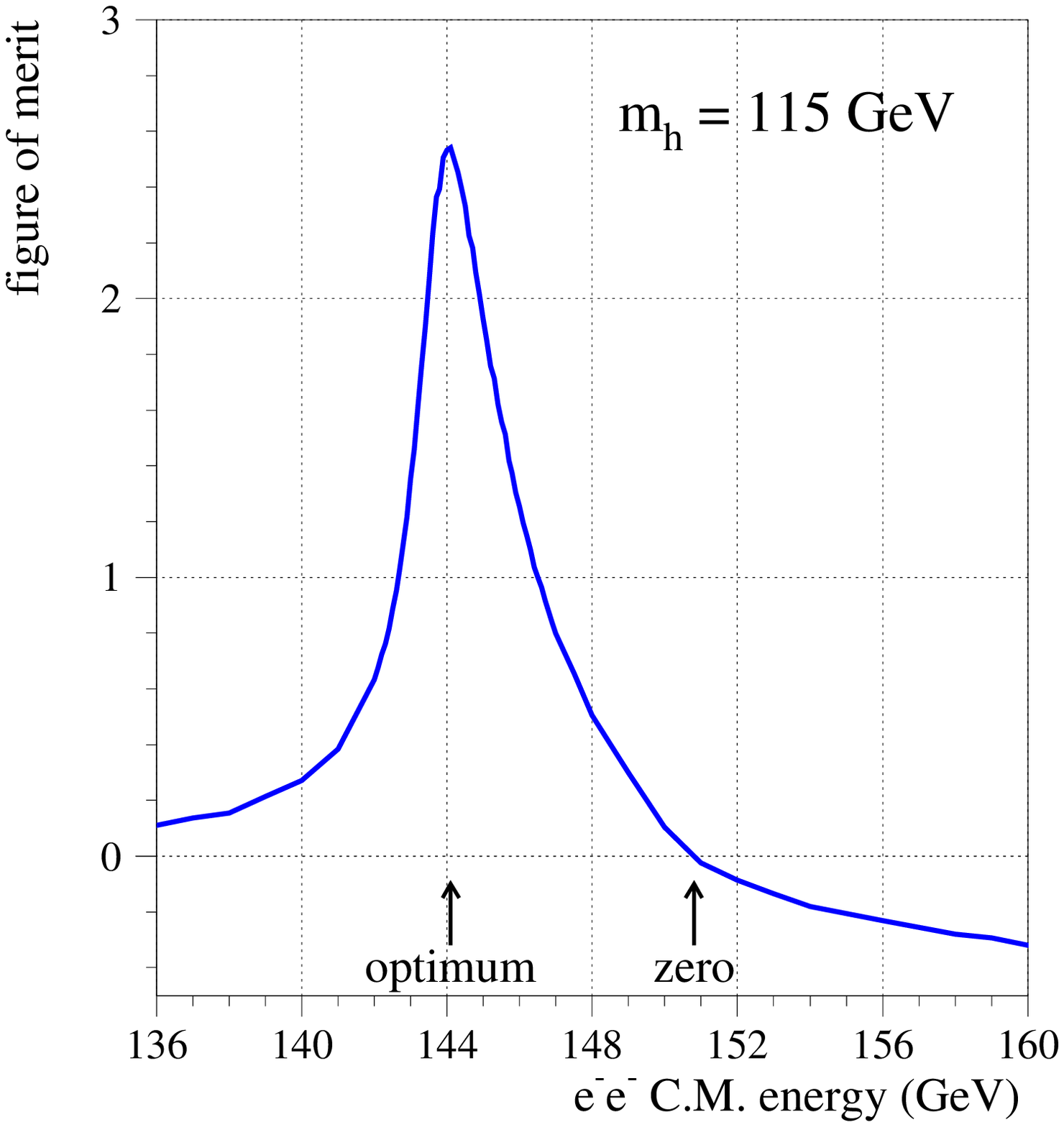,height=7cm}
\epsfig{file= 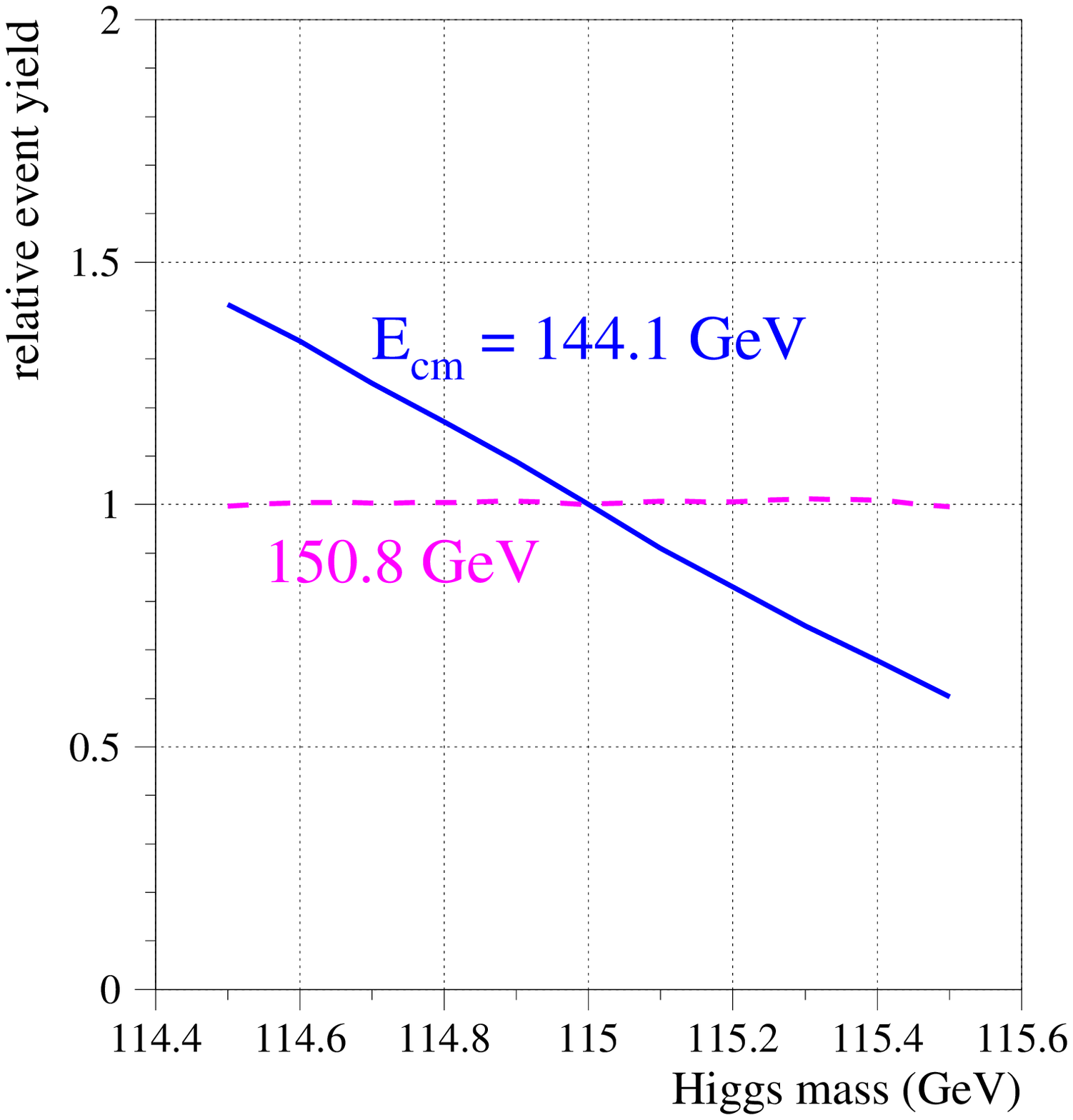,height=7cm}
\epsfig{file= 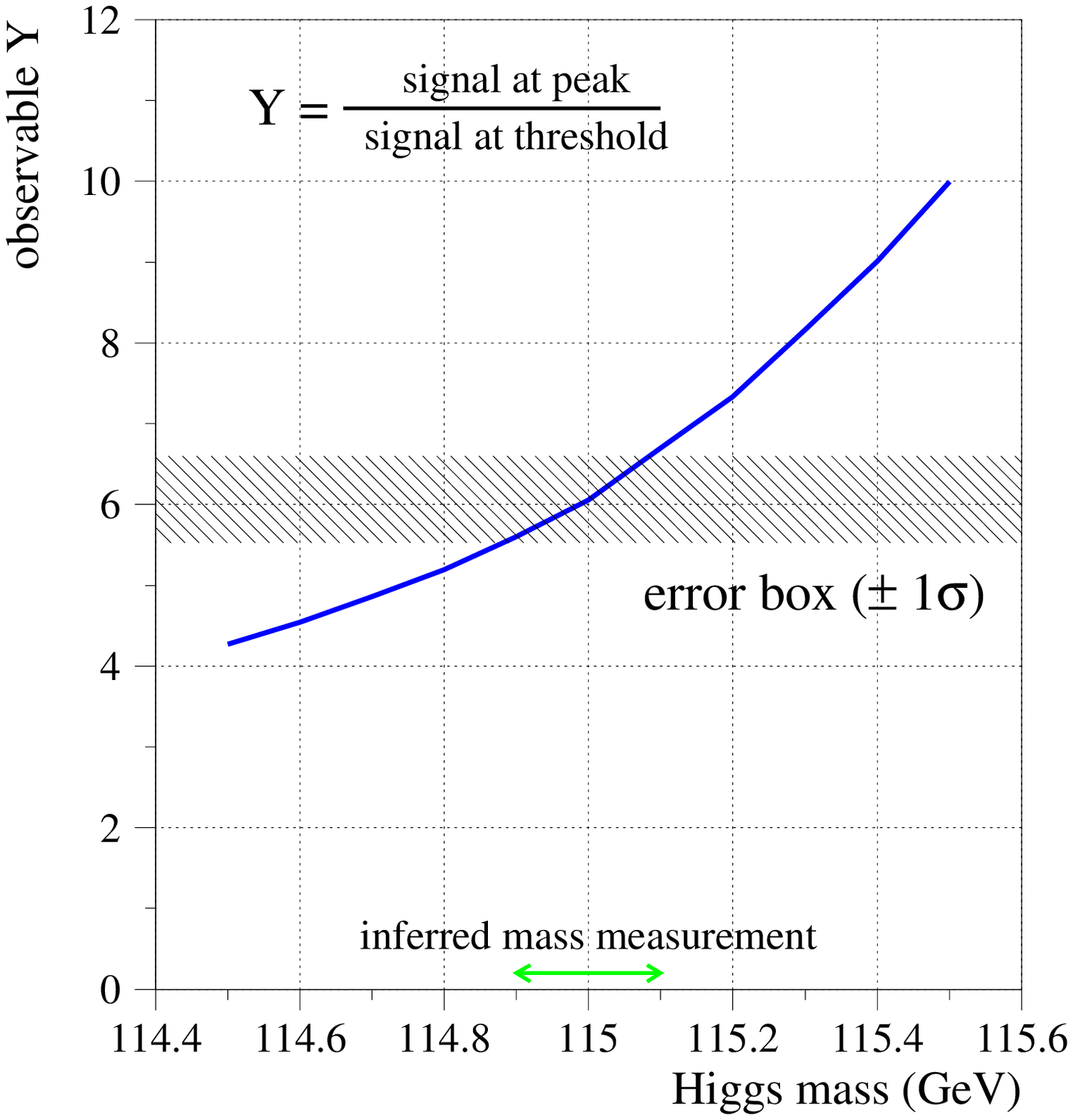,height=7cm}}
\end{center}
\caption[.]{\label{fig:scan}
\em 
(a) A figure of merit quantifying the measurement error on the mass as a function of the $e^-e^-$ centre-of-mass energy. The optimum and zero 
sensitivity points are marked.
(b) Relative yield for a 115~GeV Higgs boson at the point of optimum
sensitivity and zero sensitivity to $m_H$.  
(c) Behavior of the observable~$Y$ as a function of $m_H$, 
and the projected error.
}
\end{figure}

\begin{figure}[ht]
\begin{center}
\mbox{\epsfig{file=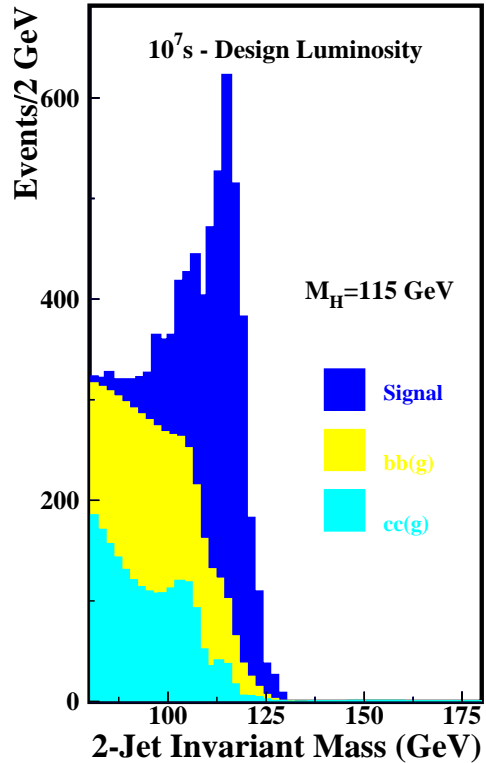,height=11cm}}
\end{center}
\caption[.]{\label{fig:hbb}      
\em
Observability of the $H \rightarrow {\bar b} b$  decay mode for
$m_H=115$~GeV, with CLICHE running so that the peak $E_{CM}(\gamma\gamma)
= 115$~GeV. 
}
\end{figure}

We have studied this possibility in the context of
CLICHE~\cite{threshold-method}, assuming that the Higgs mass is already
known to within a GeV or so, from the Tevatron, the LHC or a $e^+ 
e^-$ linear collider. 
Considered as a function of the $e^-e^-$ centre-of-mass energy, as shown in
Fig.~\ref{fig:excitation}, there is a point of optimum sensitivity to the
Higgs mass a few~GeV below the peak of the cross section. The raw number
of events at a single energy cannot be used to measure the
mass, however, because the $\gamma\gamma$ partial width cannot be assumed
known {\it a priori}. There is another point, though, close to the maximum
of the cross section, at which there is no sensitivity to the Higgs mass,
and with maximum sensitivity to $\Gamma_{\gamma\gamma}$, allowing the
separation of these two quantities. These points are illustrated in
Fig.~\ref{fig:scan}.  Furthermore, the background can be estimated using
data obtained by running below the threshold.  To estimate the 
sensitivity of the yields to $m_H$, we work with a simple 
observable based on the ratio of background-subtracted yields
at peak and at threshold:
\begin{displaymath}
 Y = \frac{N_{\mathrm{peak}} - N_{\mathrm{below}}\cdot r_p}
          {N_{\mathrm{threshold}} - N_{\mathrm{below}}\cdot r_t}
\end{displaymath}
where $N$ is the number of events in a mass window logged at the peak,
on the threshold, and below threshold, and $r_p$ and $r_t$ are scale
factors to relate the background data taken below threshold to
the expectation at peak and at threshold.  We have propagated
statistical uncertainties, and, assuming one year of data on peak,
half a year on threshold and another half below threshold, we find
$\sigma_Y / Y = 0.088$.  This translates into an error
on the inferred Higgs mass of~100~MeV.  A more refined treatment should
improve this estimate somewhat. This estimate is 
obtained using the laser and beam energies proposed for CLIC~1 and the
analysis results shown in Fig.~\ref{fig:hbb}. It is still necessary to 
investigate how  
sensitive the luminosity function  is to the shape of the luminosity curve.
It is not sensitive to the electron polarization precision.

\noindent\underline{ $ H \rightarrow {\bar b} b$:}\\
~\\
Due to the large branching ratio for $H\rightarrow\bar{b}b$ decay
for a Higgs mass $\sim 115$~GeV, this is the main channel for Higgs 
studies at CLICHE. This channel has received
the most attention and the studies are already quite
detailed~\cite{david_jack,ohgaki97,stephene}. 
Our analysis includes perturbative QCD backgrounds,
including $\gamma \gamma \rightarrow {\bar b} b(g)$ and
$\gamma \gamma \rightarrow {\bar c} c(g)$.
The ${\bar q} q$ backgrounds are suppressed by choosing like polarizations
for the colliding photons, but this suppression is not so strong when the
final states contain additional gluons.

In this analysis we used the Durham jet algorithm and imposed a 
cut at $y=0.02$ to
define the two jets.  The main cuts are: (1)  only two-jet events are
accepted, (we do not find any improvement if we include three-jet events
as well), (2) $| \cos\theta | < 0.5$, which is 50\% efficient, and (3) the
two jets are required to be back-to-back. The last cut is very important
for the background suppression, but at a significant cost in the signal.
The efficiency of this cut is 85-90\% for the 33\% of events that do not
contain neutrinos, but there is significant reduction in efficiency for
the rest of the events. We assume here that there will be a 3.5\%
$c\bar{c}$ contamination and that the $b$ tagging is 70\% efficient
for double tagging~\cite{Orange}.  The
final reconstruction efficiency is expected to be 30\%.

Recent photoproduction measurements of the $b$ cross section
at the highest energies at LEP are larger than predictions from NLO
perturbative QCD calculations. L3~\cite{L3bbar} and OPAL~\cite{OPALbbar}
report values for the cross section of $ee \rightarrow ee b\overline{b}X$
at $\sqrt{s} = 194$~GeV which are, respectively,
 3\,$\sigma$ and 2.5\,$\sigma $ larger than predicted.
Hadronic final states containing $b$ quarks are identified by detecting
leptons from their semi-leptonic decays.  The effective $W_{\gamma\gamma}$ 
energy at which these measurements are made is considerably lower than the 
one at CLICHE, and it is therefore not clear that this will affect the 
signal-to-background ratio, in particular after all selection cuts made. 
This background can be controlled by measuring off the Higgs mass peak.

The mass resolution is around 6~GeV with a jet energy resolution of
$\sigma_E=0.6 \times \sqrt{E}$. The distribution in the di-jet invariant 
mass,
$m_{jets}$, for a $m_H=115$~GeV Higgs found in this study 
with an integrated luminosity of 200~fb$^{-1}$ is shown in
Fig.~\ref{fig:hbb}. A clear signal peak can be seen above sharply falling
backgrounds. Including the three bins nearest to $m_{jets}\sim 115$~GeV,
we obtain 4952 signal events and 1100 background events. Thus, the
signal-to-background ratio is expected to be 4.5 after all cuts.

A feature which is not taken into account in these studies is the pile-up
of events from different bunch crossings.  The bunch structure of CLIC
leads to a bunch crossing every 0.67~nanoseconds. The detector components 
will generally  not have the capability to time-stamp hits with this 
resolution and will provide information integrated over several  
bunch crossings. Hence an R\&D program on the development of fast 
detectors 
will need to be pursued in the coming years (this is also the case for 
detectors at the NLC, where the separations between bunches are only a few
nanoseconds). The components most likely to determine the total number 
of bunch crossings that will be integrated over, for the studies of 
interest here,  will be the silicon pixel detectors, for which a time-stamping 
capability of  25~nsecs has been demonstrated for LHC experiments. It
can be expected that this resolution will improve perhaps to 10~nsec or even 
5~nsec within the next 10 years, based on present ideas that still need 
to be tested. 
Nevertheless, this means that of order 10 bunch crossings will contribute 
to measured events at CLICHE, which will affect the background shape of 
Fig.~\ref{fig:hbb}. This effect was not taken into acount here, but initial 
studies have shown that it deteriorates the Higgs signal only slightly.

\begin{figure}[htbp]
\begin{center}
\mbox{\epsfig{file=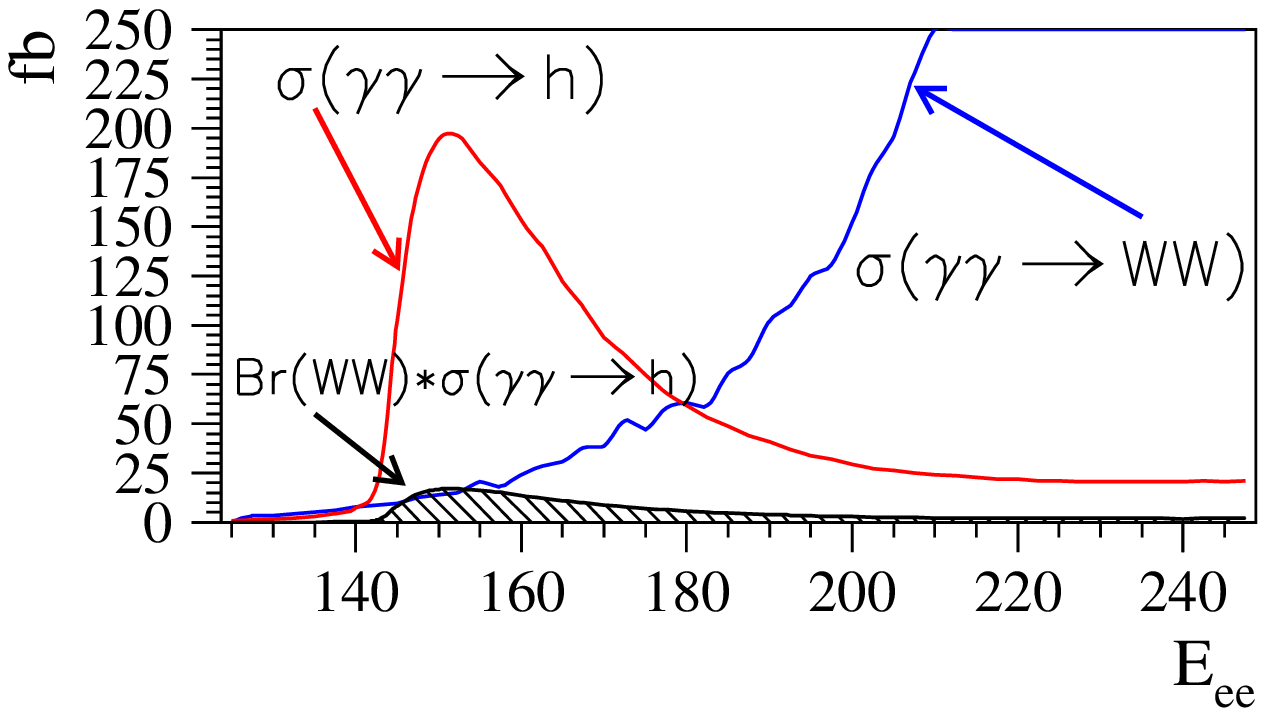,height=6.cm}}
\mbox{\epsfig{file=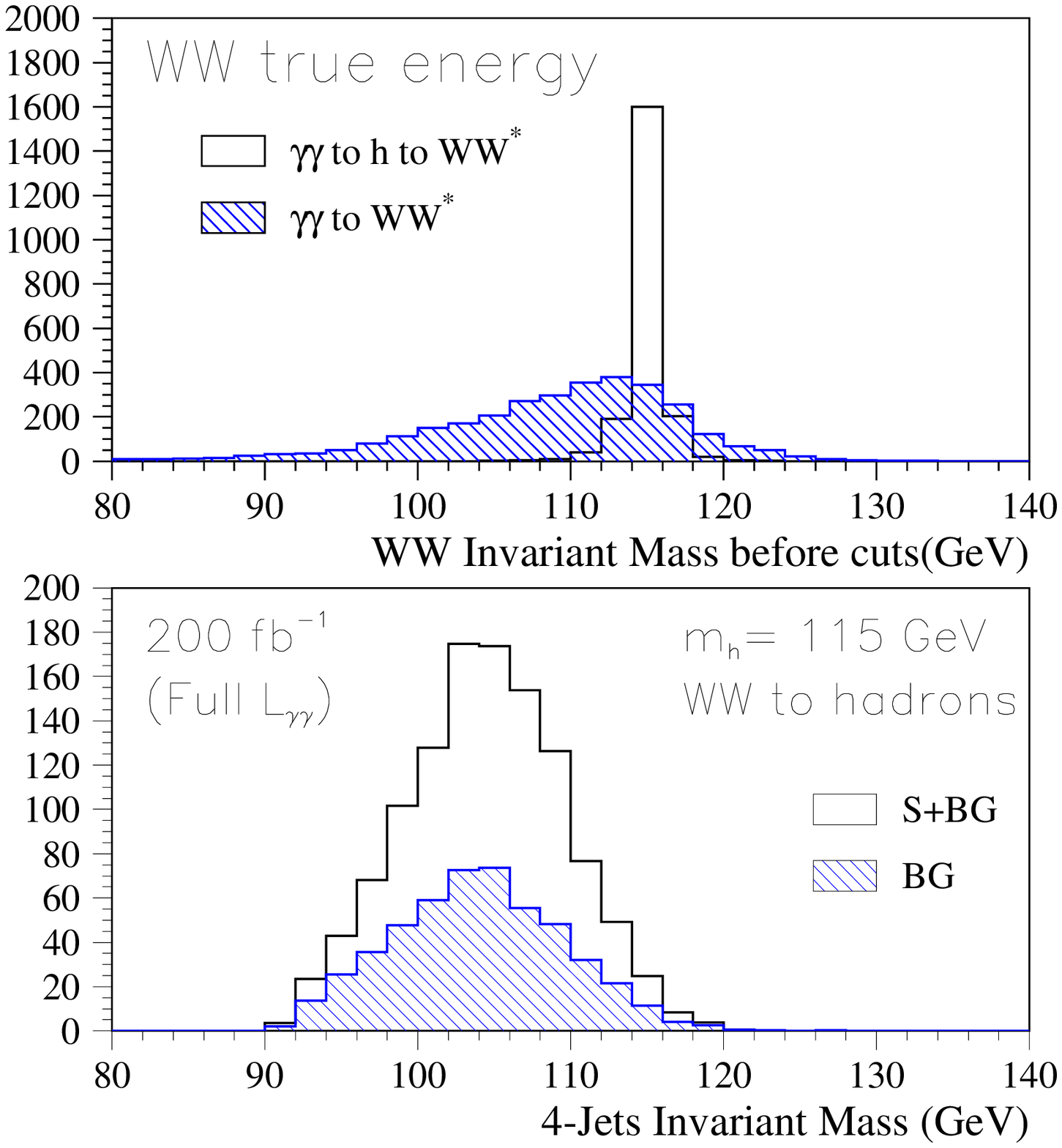,height=10.cm}}
\end{center}
\caption[.]{\label{fig:wwcross}      
\em
(a) Cross sections for $\gamma\gamma\rightarrow h$, 
$\gamma\gamma\rightarrow h \times {\cal B}r(h\to WW)$ for 
$m_H=115$~GeV and  $\gamma\gamma\rightarrow WW$
production. (b) Comparison of the  ideal invariant mass
of the $WW$ pairs  from signal and background events.
(c) Selection of the $WW$   decay mode of the Higgs boson for 
$m_H=115$~GeV, running at $E_{CM}(\gamma\gamma)=115$~GeV at CLICHE.
}
\end{figure}

\noindent\underline{ $ H \rightarrow WW $:}\\
~\\
Observation of this decay mode is extremely difficult at high-energy
$\gamma\gamma$ colliders, because of the large cross section for $W$~pair
production.  If the $\gamma\gamma$ centre-of-mass energy is below the
$W^+W^-$ threshold, however, the continuum production of $W$ pairs is
greatly reduced, allowing the observation of resonant production through a
Higgs boson.  The sharp peak in the $\gamma\gamma$ luminosity function seen in
Fig.~\ref{fig:spectra} plays a key role here.
Figure~\ref{fig:wwcross}(a) compares the cross sections for the 
continuum $W$~pair production with the Higgs resonance curve. As shown,
the cross sections for $\sigma(\gamma\gamma\to W^+W^-)$ and 
${\cal B}r(h\to W^+W^-) \times \sigma(\gamma\gamma\to h)$ are comparable,
if  $E_{CM}(e^-e^-)=150$~GeV for a  $m_H=115$~GeV. 
One significant difference between the two type of events is
the energy distribution of the $W^+W^-$ pairs, as illustrated 
in Figure~\ref{fig:wwcross}(b).

Our study is concentrated on the hadronic decays of the $W$ pairs,
applying several kinematic cuts. One pair of jets must reconstruct to the
$W$~mass, while the other pair is required to saturate the remaining phase
space. This cuts allows us not only to reduce the  $W^+W^-$ pairs 
to those with energy similar to those  produced in Higgs events, but 
also to reject
 any possible $\gamma\gamma \to qq(g)$ background.  There must be at least
four jets in the event and  the jet reconstruction efficiency is assumed to 
be 100\%.  In constrast to the $h\to b\bar{b}$ analysis, here we are imposing 
a $y=0.003$ cut  in the Durham algorithm used in the jet reconstruction.
In addition, the transverse momentum is required  to be smaller than 0.1. 
After these cuts we have a 29\% reconstruction efficiency.
A comparison of the signal and the  background after cuts is given in
Fig.~\ref{fig:wwcross}(c), which corresponds to a signal-to-background ratio
of 1.3, and the statistical precision in the
signal rate measurement is expected to be 5\%.

The other event topologies (two leptons and missing energy, or one lepton,
missing energy and jets) remain to be studied.  Techniques similar to those
described in \cite{dittmar} may be used. We also believe that the
decay $H \rightarrow ZZ, Z\gamma$ might be interesting, despite their
relatively small branching ratios.

\noindent\underline{ $ H \rightarrow \gamma\gamma $:}\\
~\\
In almost any phenomenological context, the decay $H \rightarrow
\gamma\gamma$ is a very rare one. However,
the number of Higgs events is large at a $\gamma\gamma$~collider, 
so an interesting number of $H \to \gamma\gamma$
events would be produced.  Furthermore, the backgrounds are expected
to be quite small, below 2~fb~\cite{jikia_gg}, 
 since there is no tree-level coupling of photons,
and the box-mediated processes are peaked very sharply in the
forward direction. A  complete background study has not yet been made, but 
initial estimates indicate that a clear peak
in the $\gamma\gamma$~mass distribution should be observable, 
and we assume here that the background error would be negigible.

\begin{figure}[hb]
\begin{center}
\mbox{\epsfig{file=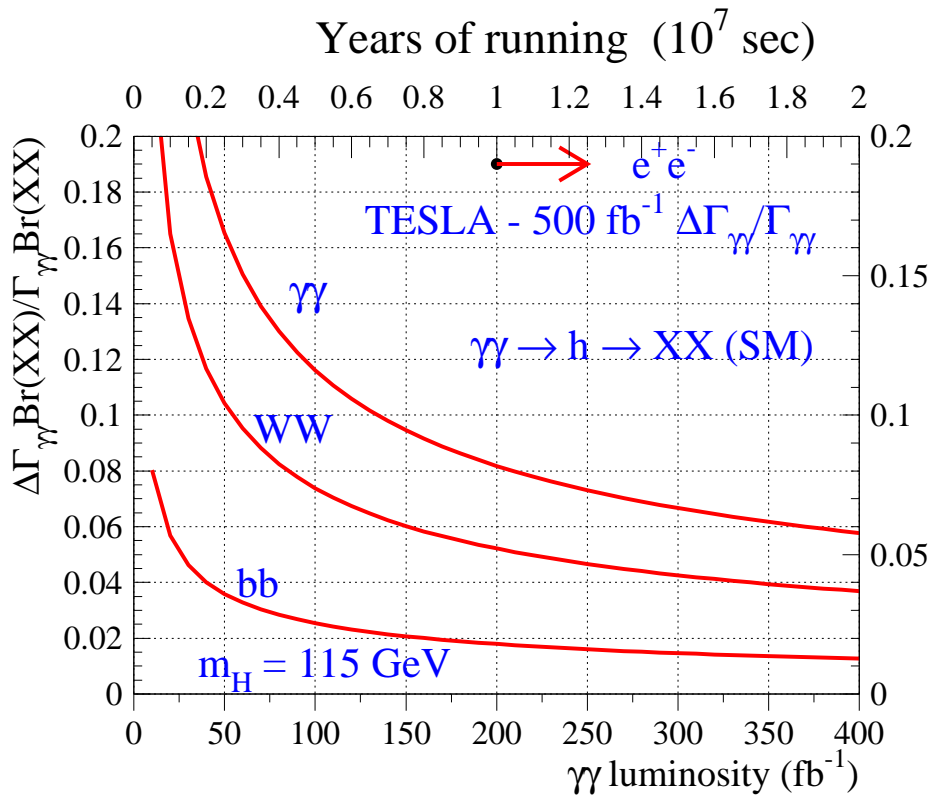,height=11.00cm}}
\end{center}
\caption[.]{\label{fig:ggtogg}      
\em
The expected precision in the ${H \rightarrow \gamma\gamma}$ decay
width from direct measurements
of $H\rightarrow\gamma\gamma$ for $m_H = 115$~GeV. The precision is
less than in the equivalent measurement of $H\rightarrow WW, \bar{b}b$,
but this observable is unique to a low-energy $\gamma\gamma$ collider
like CLICHE. 
}
\end{figure}

The number of events produced in this channel is proportional
to ${\Gamma_{\gamma\gamma}^2/\Gamma_{\mathrm{total}}}$. The quadratic
dependence is interesting, because if $\Gamma_{\mathrm{total}}$ could
be measured elsewhere, a small error on $\Gamma_{\gamma\gamma}$ would
be obtained. Similary, if $\Gamma_{\gamma\gamma}$ is measured elsewhere, 
a small error $\Gamma_{\mathrm{total}}$  could be obtained.
In Fig.~\ref{fig:ggtogg}, we can see that a 10\% measurement of
${\Gamma_{\gamma\gamma}^2/\Gamma_{\mathrm{total}}}$ can be made with 
less than a year of data taking.

The cleanliness of these events and good energy resolution in the 
electromagnetic calorimeter would allow for an independent measurement 
of the Higgs mass. Assuming that the calorimeter energy scales can
be sufficiently well calibrated, a resolution better than $100$~MeV
can be expected.

\noindent\underline{Other Channels:}\\
~\\
A preliminary assessment of the $ H \rightarrow {\bar c} c$ channel at a
$\gamma \gamma$ collider is not very encouraging: relative to the ${\bar b}
b$ mode, the ${\bar c} c$ signal is suppressed by factor of $(m_c/m_b)^2
\sim 1/10$, and the background is enhanced by a factor $(Q_c/Q_b)^4 = 16$.  
There is similar pessimism concerning the observability of $ H \rightarrow
g g$ and $ H \rightarrow \tau^+ \tau^-$.

\noindent\underline{Combining Channels:}\\
~\\
A good measurement of the two-photon partial width, $\Gamma_{\gamma\gamma}$, 
is very important as it receives direct contributions from all charged 
massive particles, and there is no tree-level contribution.  Since the
Higgs production cross section is proportional to  $\Gamma_{\gamma\gamma}$,
the measurement of any yield provides information, in principle, on
$\Gamma_{\gamma\gamma}$.

From the study of the channel $H \rightarrow \bar{b}b$ we estimate a
precision of 2.0\% on the quantity $\Gamma_{\gamma\gamma}\, \times\, {\cal
B}r(H\rightarrow\bar{b}b)$. The branching ratio ${\cal
B}r(H\rightarrow\bar{b}b)$ cannot be measured directly at a $\gamma\gamma$
collider, but there are methods for measuring it to about 1.5\% at an
$e^+e^-$ linear collider~\cite{nlc_report,TESLA_TDR}.  By combining the
information from both colliders, $\Gamma_{\gamma\gamma}$ can be inferred
to a precision of 2.2\%, to be contrasted with a 19\% measurement from
500~fb$^{-1}$ at TESLA (without the $\gamma \gamma$ option), for example.

\begin{figure}[ht]
\begin{center}
\mbox{\epsfig{file=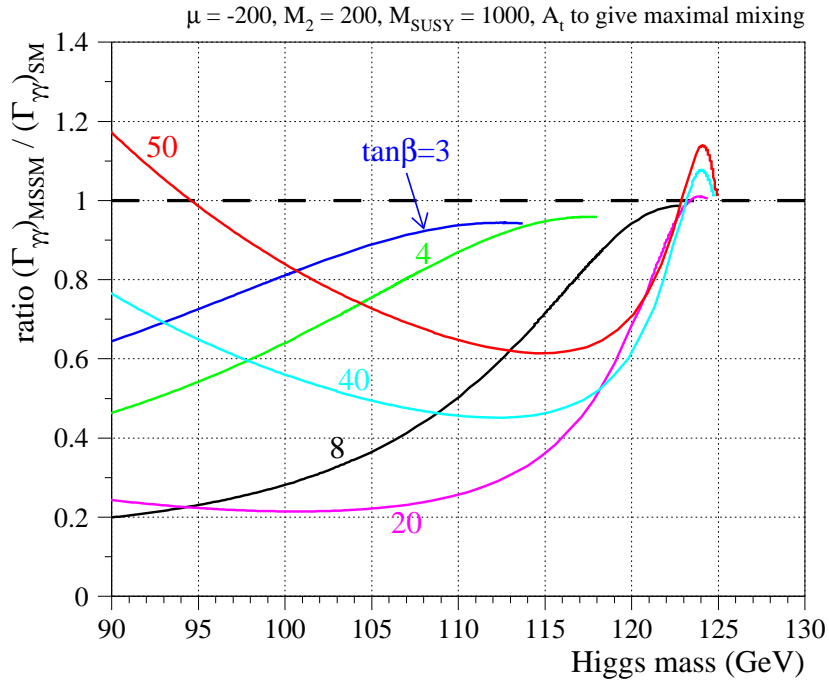,height=9cm}}
\end{center}
\caption[.]{\label{fig:susygaga}\it
The ratio of the partial width $\Gamma_{\gamma\gamma}$ in the MSSM to that in 
the SM~\cite{hdecay}.  The `maximal mixing'~\cite{higgscorr}
scenario has been chosen for this illustrative plot; this scenario tends to 
give the largest Higgs masses for a given $\tan\beta$ and $M_A$.
These curves do not necessarily map out the largest possible
variation in this ratio, nor have they been constrained by
negative searches for Higgs bosons.
The ratio of cross sections follows the same curves.
}
\end{figure}

As shown in Fig.~\ref{fig:susygaga}, the $\Gamma_{\gamma\gamma}$
partial width in the MSSM can deviate a lot from the SM value; even
for a fixed Higgs mass of 115~GeV, for example, a factor of two
variation is possible.  Note that this impacts the cross section and
hence precision on the inferred value for $\Gamma_{\gamma\gamma}$.
Nonetheless, it is clear that a precision of 2--4\% would be
very discriminating in the context of an unconstrained MSSM.

Alternatively, one could take the measurement of the yield in the
$\bar{b}b$ channel at a $\gamma\gamma$~collider and combine it with
measurements of the Higgs branching ratios to $\bar{b}b$ and $\gamma\gamma$
from an $e^+e^-$ linear collider, 
and extract an indirect value for the total width:
\begin{equation}
\label{eq:totalwidth}
  \Gamma_{\mathrm{total}} = \frac{
    \lbrace \Gamma_{\gamma\gamma} \times {\cal B}r(H\rightarrow\bar{b}b) \rbrace
                              }  {
   \lbrace {\cal B}r(H\rightarrow\gamma\gamma) \rbrace \, \times \,
   \lbrace {\cal B}r(H\rightarrow\bar{b}b) \rbrace
                              } .
\end{equation}
According to~\cite{battaglia}, the anticipated precision
on ${\cal B}r(H\rightarrow\bar{b}b)$ and ${\cal B}r(H\rightarrow\gamma\gamma)$ 
are 2.4\% and 19\%, respectively.  Combined with a precision of 2\%
on the numerator, this leads to a precision of about 20\% for
$\Gamma_{\mathrm{total}}$.  

A better measurement of $\Gamma_{\mathrm{total}}$ can be obtained by
combining the 2\% and 8\% measurements of $H \rightarrow \bar{b}b$ and
$H \rightarrow \gamma\gamma$ with the expected  1.5\% measurement of the 
${\cal B}r(H\rightarrow\bar{b}b)$  from  the 
$e^+e^-$ linear collider~\cite{TESLA_TDR,nlc_report}.
In this case a 9\% measurement of the $\Gamma_{\mathrm{total}}$ 
can be inferred, compared to  the  4-6\%  measurement from
500~fb$^{-1}$ at TESLA~\cite{TESLA_TDR}.

\begin{table}[t]
\caption{\it The statistical errors on selected decay modes of
a 115~GeV Higgs boson in the Standard Model. The $\gamma\gamma\to h$
cross section for the full~(peak)  ${\cal L_{\gamma\gamma}}$ given in 
Table~\ref{t:gamma2} is 112~(624)~fb.  The expected yield for 
200~(36)~fb$^{-1}$ is 22,400  Higgs particles.}
\label{table:decay}
\begin{center}
\begin{tabular}{lccccc}
\hline
decay mode\,\,\,\,\,  &\,\,\,\,\, raw events/year\,\,\,\,\, &\,\,\,\,\, S/B\,\,\,\,\,  & \,\,\,\,\,\,\,\,\,\,$\epsilon_{sel}$\,\,\,\,\, &\,\,\,\,\,\,\,\,\,\, ${\cal B}r$\,\,\,\,\, &\,\,\,\,\, $\Delta \Gamma_{\gamma\gamma} {\cal B}r/\Gamma_{\gamma\gamma}{\cal B}r$ \\
\hline
${\bar b} b$      & 16509         & 4.5  &   0.30 &  73.7\% & 2\% \\
\hline
$W^+ W^-$         & 1971            & 1.3  &  0.29  &  8.8\% & 5\%\\
\hline
$\gamma\gamma$    &  45             & ---  &  0.70  &  0.2\% & 8\%\\
\hline\\
\end{tabular}
\end{center}
\end{table}

\noindent\underline{CP}\\
~\\
A measurement that might be unique for $\gamma\gamma$ collider 
experiments like CLICHE could be that of the CP
properties of the $ H \rightarrow \gamma \gamma$ vertex~\cite{gonaris},  
which can be 
measured by colliding photons with orthogonal linear polarizations, which 
define initial states of definite CP. In the Standard Model, the $ H
\rightarrow \gamma \gamma$ vertex has only a CP-even part, but in extended
models, such as supersymmetry, there may also be a CP-odd part, which
could provide an interesting window on the mystery of CP violation. The CP
properties of the $ H \rightarrow \gamma \gamma$ vertex are in principle
distinct from those of other Higgs vertices, and hence have independent
interest.  This measurement would require higher-energy electrons
and producing the photons using lasers of longer wavelength in order to
reduce the interference and obtain a state with CP better 
defined~\cite{hagiwara,david_jack}.

Another window on the CP nature of the Higgs boson is provided by angular
distributions in the $H \rightarrow W^+W^-$ channel~\cite{snowmass}. 
This channel is
useful because information can be obtained even in the absence of
linearly-polarized photon beams. A rough estimate gives $\delta A/A \sim
5$\% or better from the measurements that could be made at CLICHE. 


\noindent\underline{Summary:}\\
~\\
We have briefly discussed measurement possibilities in the
$\bar{b}b$, $W^+W^-$ and $\gamma\gamma$ channels.
The observabilities and statistical errors of the products $\sigma (\gamma
\gamma \rightarrow H) \times {\cal B}r(H \rightarrow X)$ for each of these
decay modes are sumarized in Table~\ref{table:decay}. 
Preliminary studies indicate that the systematic errors, {\em e.g.}, 
those due to the luminosity and polarization uncertainties,
could be controlled to the same level. Thus it seems possible to measure
$\sigma (\gamma \gamma \rightarrow H \rightarrow {\bar b} b, WW, \gamma
\gamma)$ with overall precisions of 2, 5, 8\%, respectively
(see Fig.~\ref{fig:ggtogg}).

In addition, the Higgs mass can be measured three ways (fitting the peaks
in the $\bar{b}b$ and $\gamma\gamma$ mass distributions, and by the
threshold method), and the partial width $\Gamma_{\gamma\gamma}$ can be
extracted on the basis of a measurement of ${\cal
B}r(H\rightarrow\bar{b}b)$ from an $e^+e^-$ collider to very good accuracy,
not matched by any other method.  Finally, possible CP asymmetries
could be measured with a precision of about 5\%.


\subsubsection{Complementarity with other Machines}

Each of the combinations $\Gamma ( H \rightarrow \gamma \gamma ) \times
{\cal B}r(H \rightarrow {\bar b} b, WW, \gamma \gamma)$ measurable at
CLICHE is distinct from the quantities observable previously at the
Tevatron: $\Gamma ( H \rightarrow W^+ W^- ) \times {\cal B}r(H \rightarrow
{\bar b} b)$, and the LHC: $\Gamma ( H \rightarrow g g ) \times {\cal B}r
(H \rightarrow \gamma \gamma)$ and $\Gamma ( H \rightarrow {\bar t} t ) 
\times {\cal B}r (H \rightarrow {\bar b} b)$. It is estimated that the
Tevatron observable could be measured with a precision of 20\% and the LHC
observables with precisions $\sim 7, 10$\%, respectively.  The CLICHE
measurement of $\Gamma ( H \rightarrow \gamma \gamma ) \times {\cal B}r (H
\rightarrow {\bar b} b)$ would therefore be complementary to, and of
higher accuracy than, these previous measurements, whilst the other CLICHE
measurements would also be competitive. 

At an $e^+ e^-$ collider, the Higgs bosons to be used in the the 
branching-ratio 
measurements are observed in the Higgsstrahlung production process
$e^+e^- \rightarrow H + Z$, with the $Z \rightarrow \ell^+\ell^-$ fully
reconstructed. A data set of 500~fb$^{-1}$ with 
$E_{CM}(e^+e^-)$~=~350~GeV\,(500~GeV), corresponding to one to three years of 
TESLA (NLC), provides a
sample of 2200 to 3500 Higgs particles produced in the
$H + (Z \rightarrow \ell^+\ell^-)$ channel, for $m_H$ = 120 to 200~GeV.
The raw number of Higgs particles to be produced in one of these machines
is around 20,000.
An $e^+ e^-$ collider with energy $\sim 500$~GeV would be able to measure
with high precision $\Gamma (H \rightarrow ZZ)$ and all the dominant decay
branching ratios $H \rightarrow {\bar b} b, {\bar c} c, gg, \tau^+ \tau^-$
shown in Fig.~\ref{fig:decaymodes}. Moreover, such an $e^+ e^-$ collider
could also be configured as a $\gamma \gamma $ collider, if it is equipped
with a laser backscattering facility. This possibility is envisaged in 
the current designs of the JLC~\cite{jlc}, NLC~\cite{nlc_report} and
TESLA\cite{teslatdr}, but may not be scheduled for the initial phases of 
these machines.

At CLICHE, one could have 150~fb$^{-1}$ in a year 
with $E_{CM}(e^-e^-)$~=~150~GeV and
a $\gamma\gamma$ centre-of-mass energy peaked at 115~GeV.
We recall that a 115~GeV Higgs would be produced as an $s$-channel 
resonance, and that the event yield is estimated to be around   
22,000 per year. Similar yields could be expected for $m_H \le 125$~GeV 
(see Fig.~\ref{fig:excitation}(b)), if the extra energy was made 
available.

\subsubsection{Discriminating between Higgs Models using CLICHE 
Measurements}

Since the $ H \rightarrow \gamma \gamma$ vertex is due to loop diagrams,
it is sensitive to physics beyond the direct physics reach of CLICHE. For
example, a 3\% measurement of $ \Gamma (H \rightarrow \gamma \gamma)$
would provide indirectly a 6\% measurement of $\Gamma (H \rightarrow {\bar
t} t)$, in the absence of new physics~\cite{ggttbar}. However,  
supersymmetry is a prime example of
possible new physics that could influence the $ H \rightarrow \gamma
\gamma$ vertex, as seen in Fig.~\ref{fig:susygaga}.

As an example how the precision achievable with CLICHE could help
distinguish between models of Higgs bosons, in the following we compare
Higgs production in $\gamma \gamma$ collisions in the Standard Model and
its minimal supersymmetric extension, the MSSM. We do this by calculating
the product of the production cross section and branching ratio for the
lightest MSSM Higgs boson, $h$,  normalized to the corresponding Standard Model
value, with $M_H$ set to the mass of the lightest MSSM Higgs, $M_h$:
\beq
R_{b,W,\ga} := \frac{
\left[ \Gamma(\gagah) \times
       {\cal B}r(\hbb, WW^*,\ga\ga) \right]_{\rm MSSM} }
{ \left[ \Gamma(\gagaH) \times
          {\cal B}r(\Hbb, WW^*,\ga\ga) \right]_{\rm SM}}~.
\eeq
Regions of the MSSM parameter space with strong deviations from the 
Standard Model are identified in the following. 
Strong suppression of these ratios could occur in problematic corners of 
the MSSM parameter space, where a specific decay channel or even MSSM 
Higgs production itself is unaccessible. 
The evaluation of the Higgs boson sector has been performed with the codes
{\tt FeynHiggs}~\cite{feynhiggs}, based on \cite{higgscorr},
and {\tt Hdecay}~\cite{hdecay}.

\begin{figure}[htbp]
\begin{center}
\resizebox{\textwidth}{!}
{
\epsfig{file= 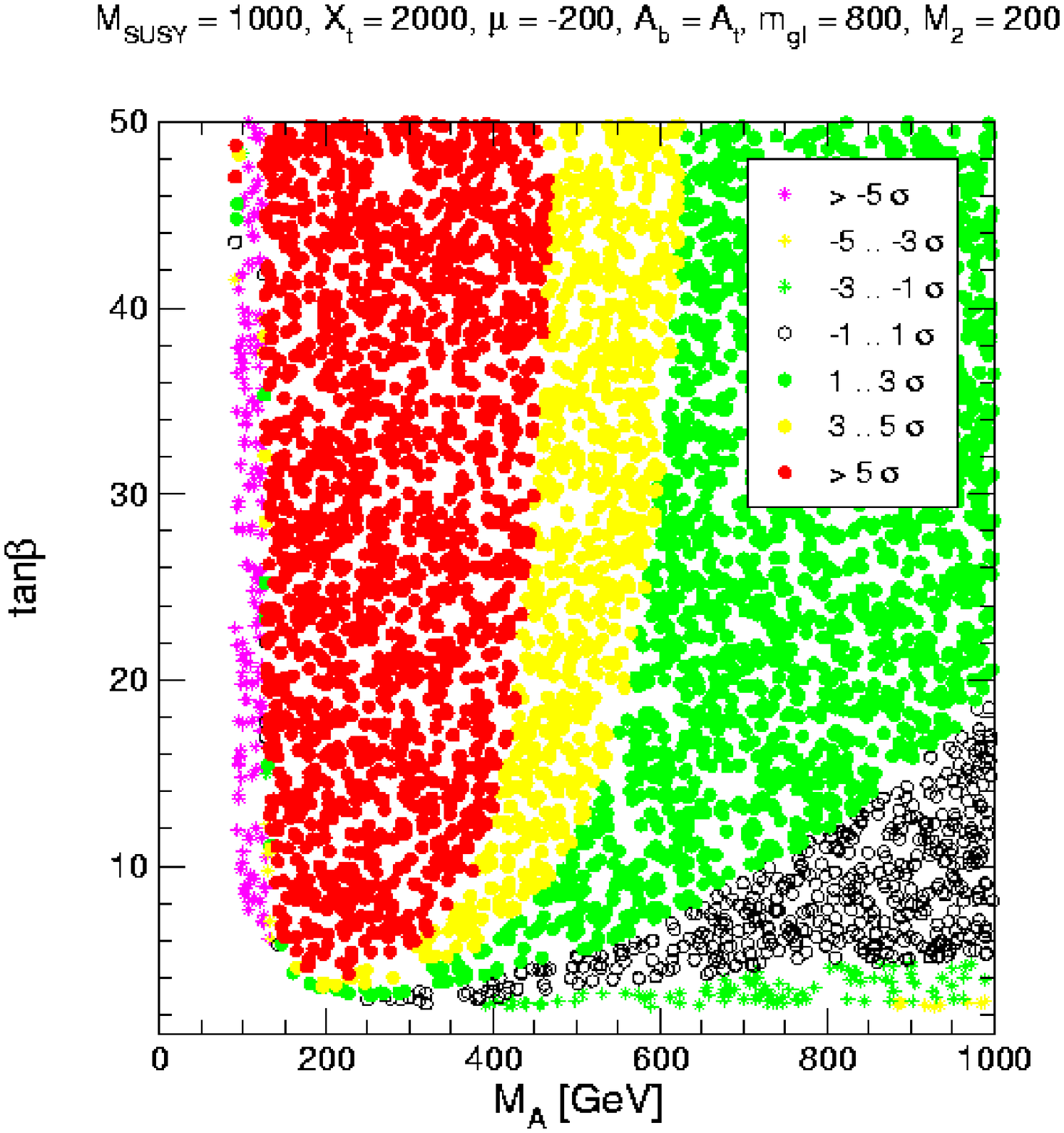,height=7cm}
\epsfig{file= 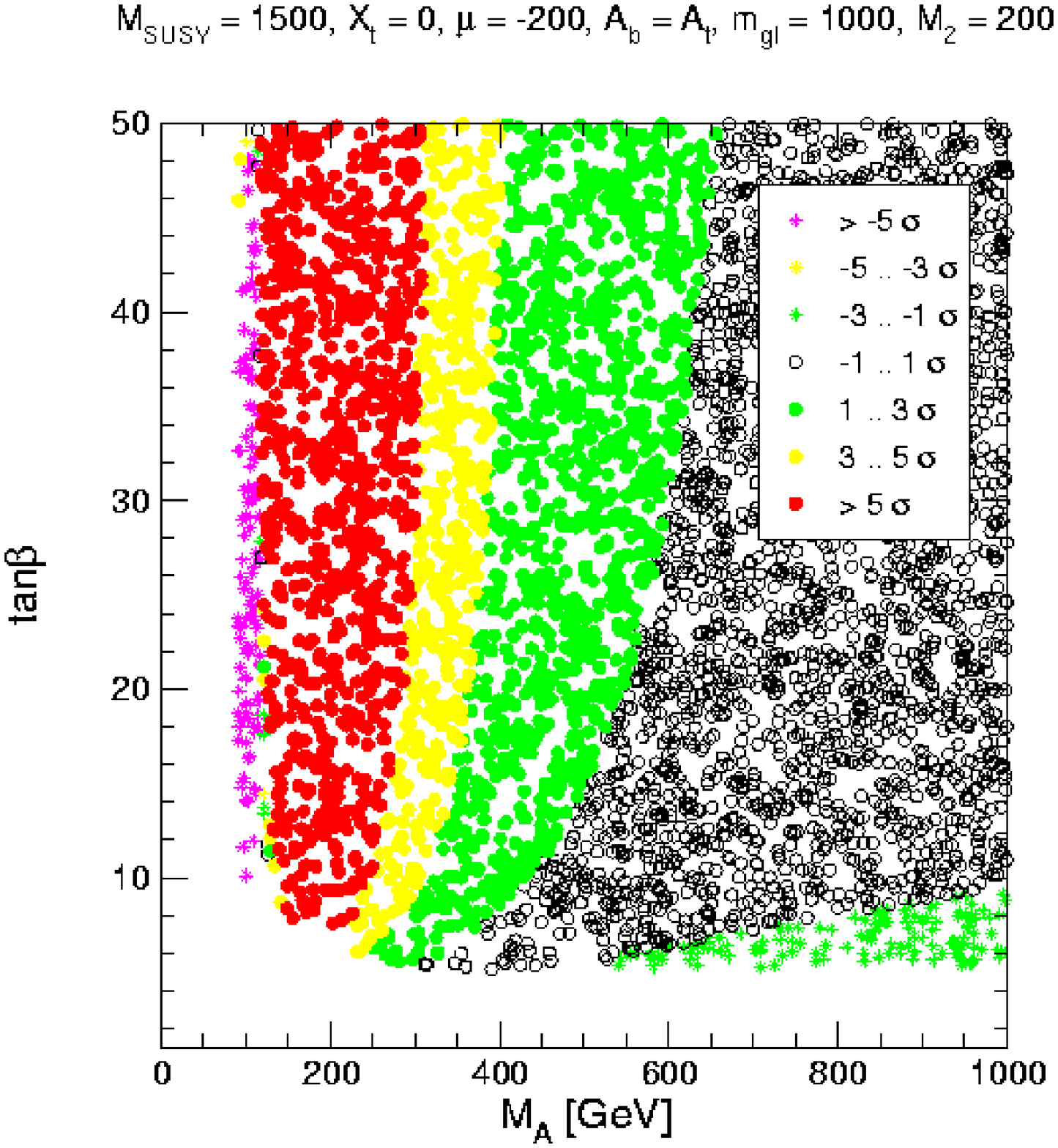,height=7cm}
\epsfig{file= 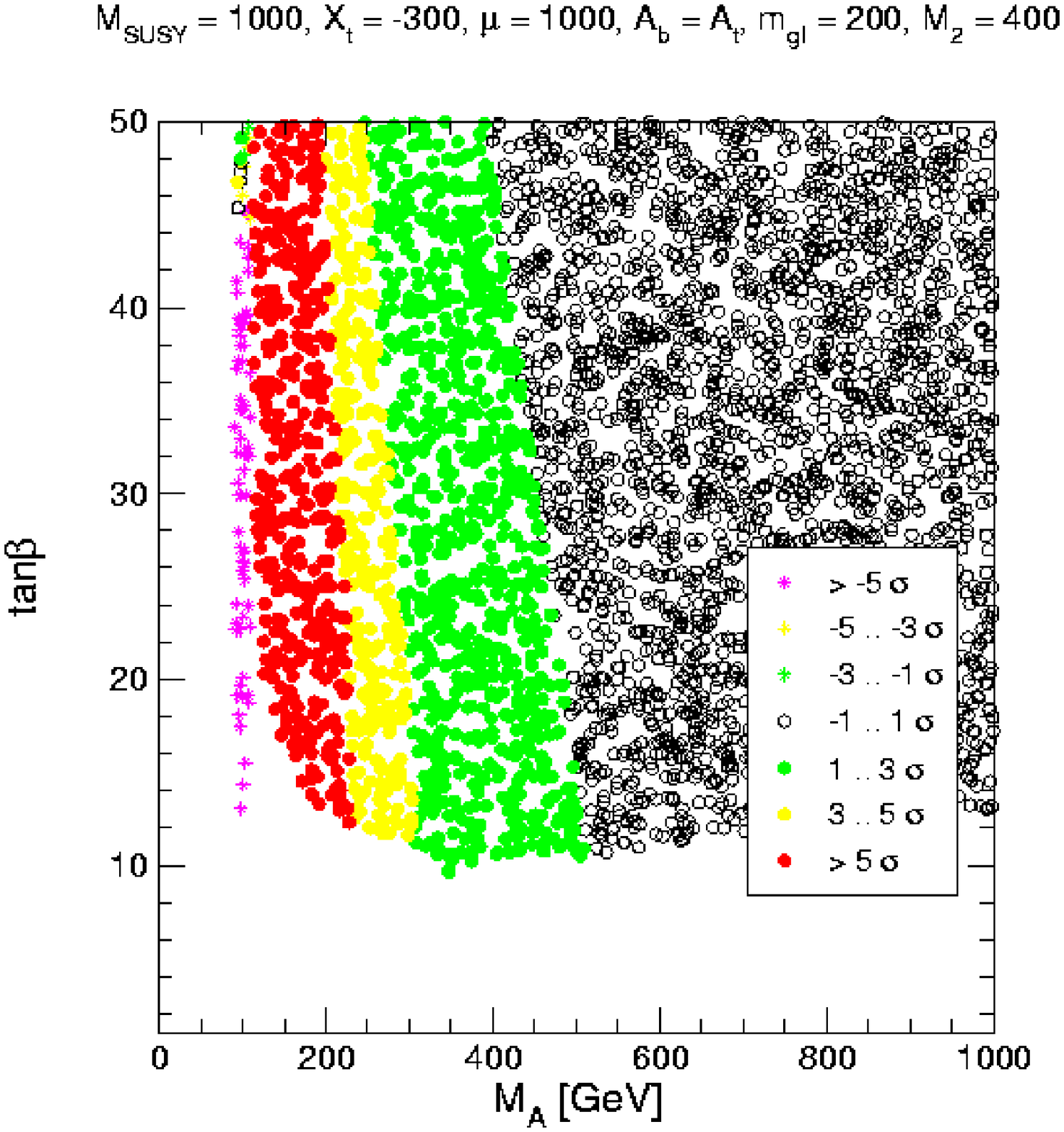,height=7cm}
}
\end{center} 
\caption[.]{\label{fig:mssmhbb}\it
The ratio $R_b$ is shown in the $\MA, \tb$ plane for the three benchmark
scenarios described in the text. The observable deviation from the 
Standard Model is indicated for a 2\% experimental precision in $R_b$.
}
\end{figure}

\begin{figure}[htbp]
\begin{center}\resizebox{\textwidth}{!}
{
\epsfig{file= 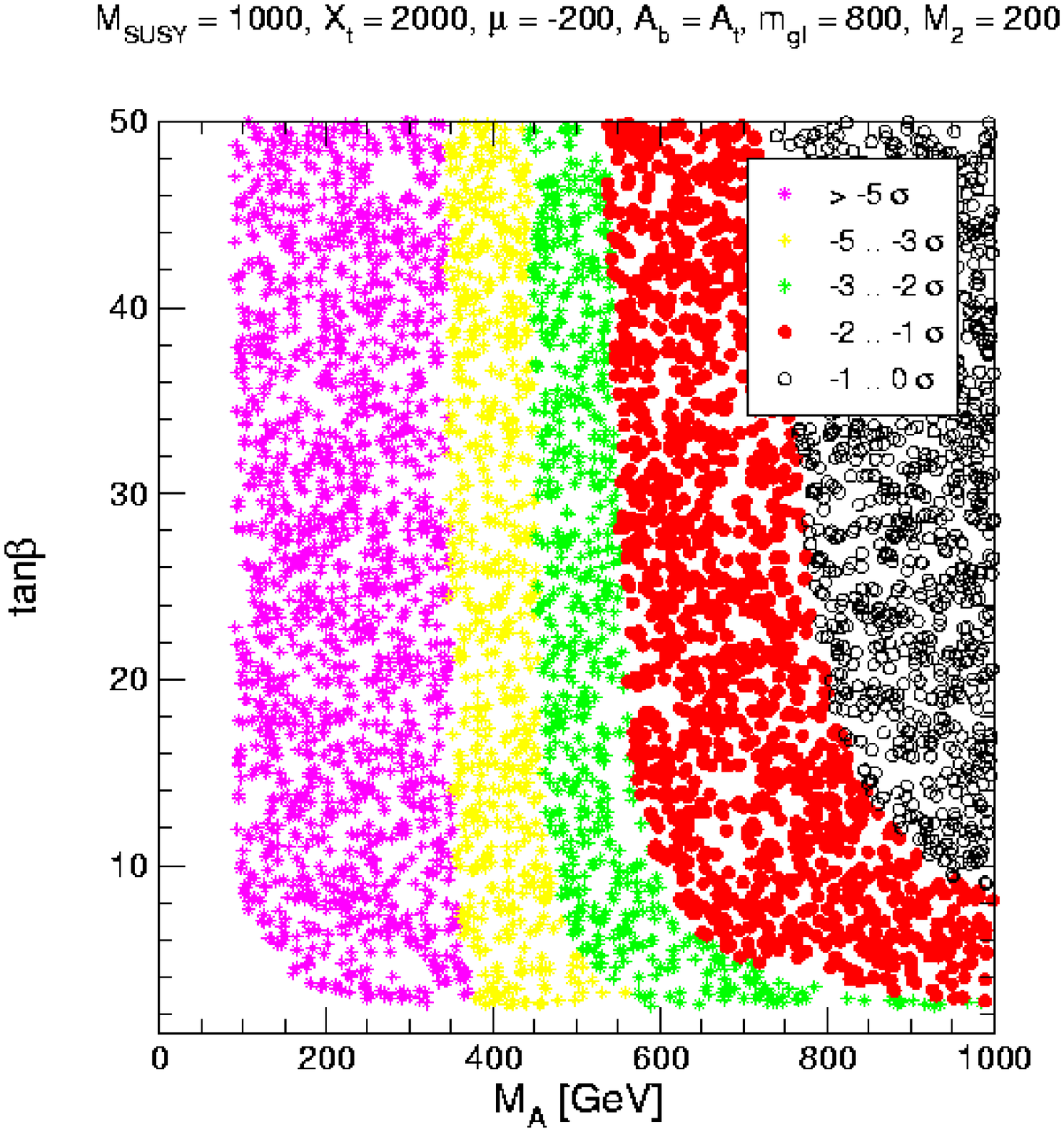,height=7cm}
\epsfig{file= 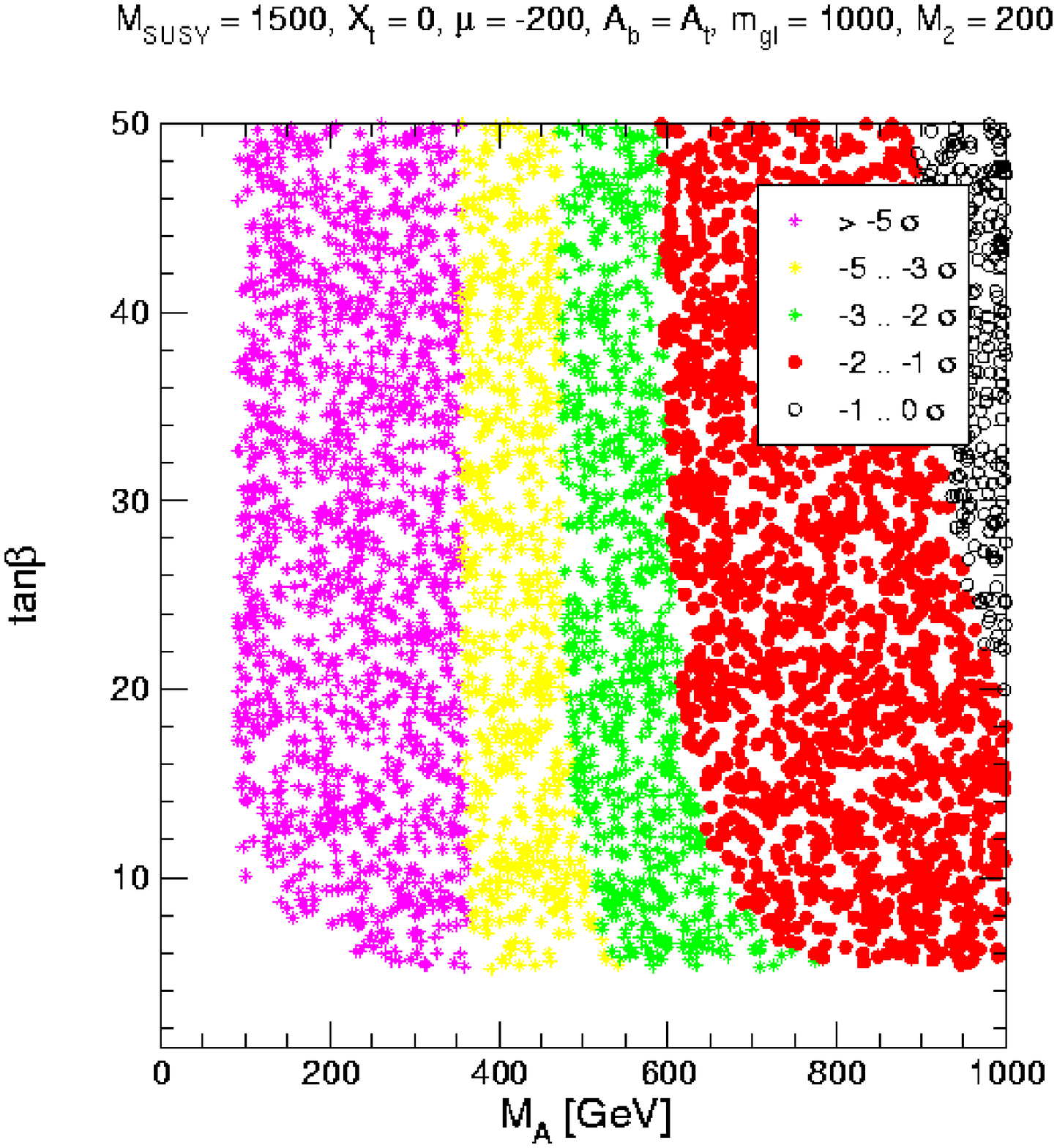,height=7cm}
\epsfig{file= 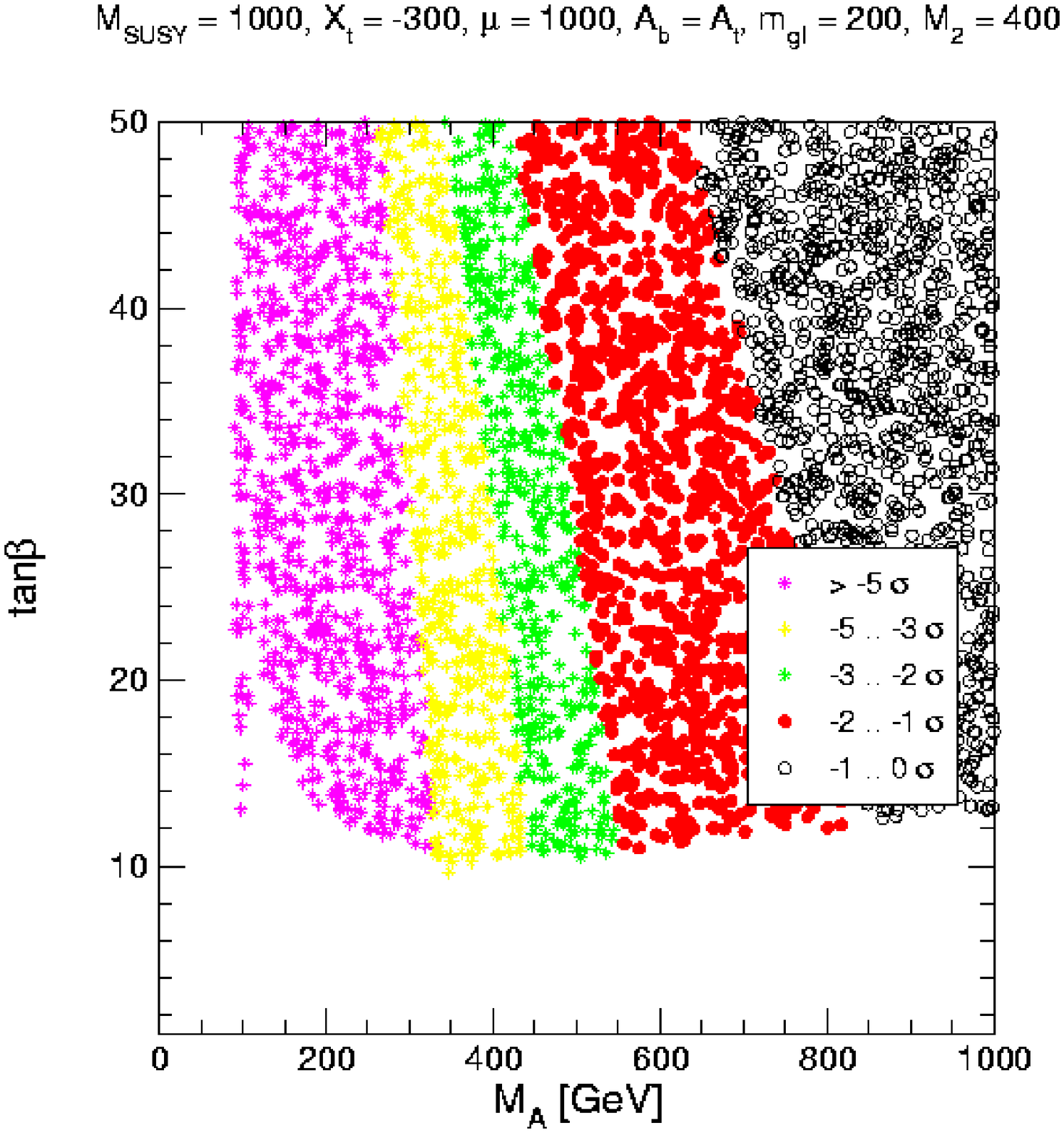,height=7cm}
}
\end{center}
\caption[.]{\label{fig:mssmhww}\it
The ratio $R_W$ is shown in the $\MA, \tb$ plane for the three benchmark
scenarios described in the text. The observable deviation from the SM
is indicated for a 5\% experimental precision in $R_W$.
}   
\end{figure}

\begin{figure}[htbp]
\begin{center}\resizebox{\textwidth}{!}
{
\epsfig{file= 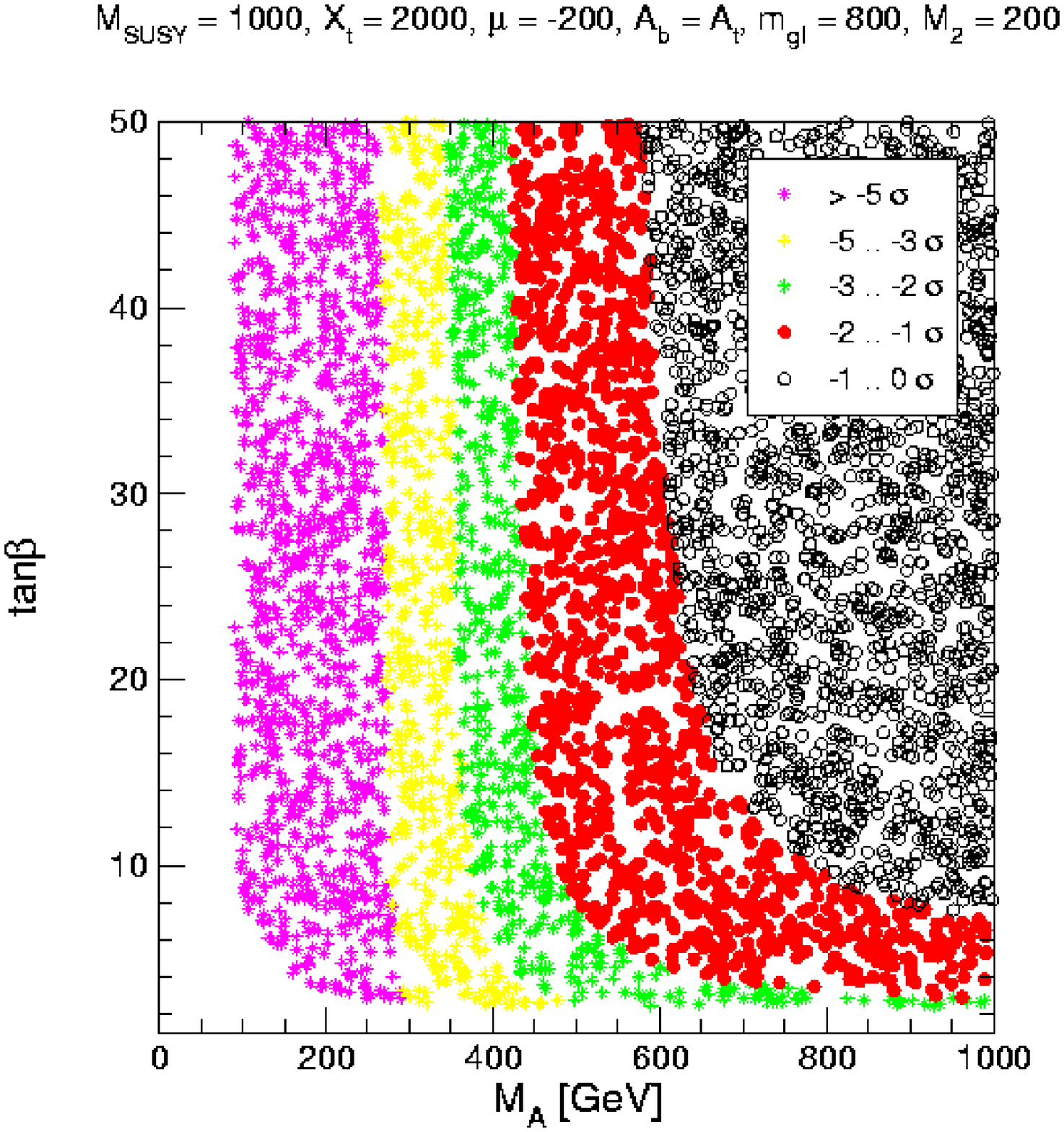,height=7cm}
\epsfig{file= 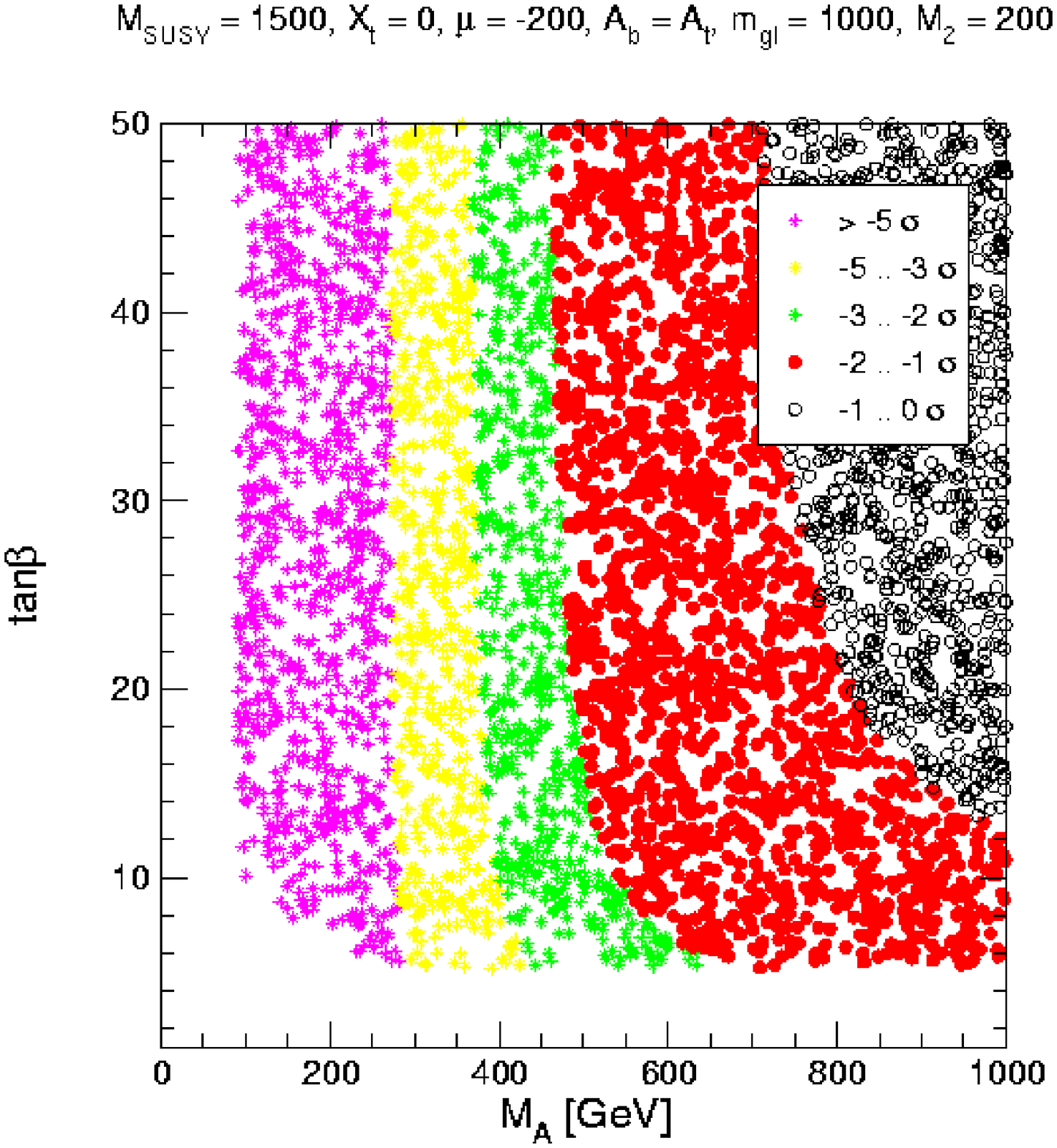,height=7cm}
\epsfig{file= 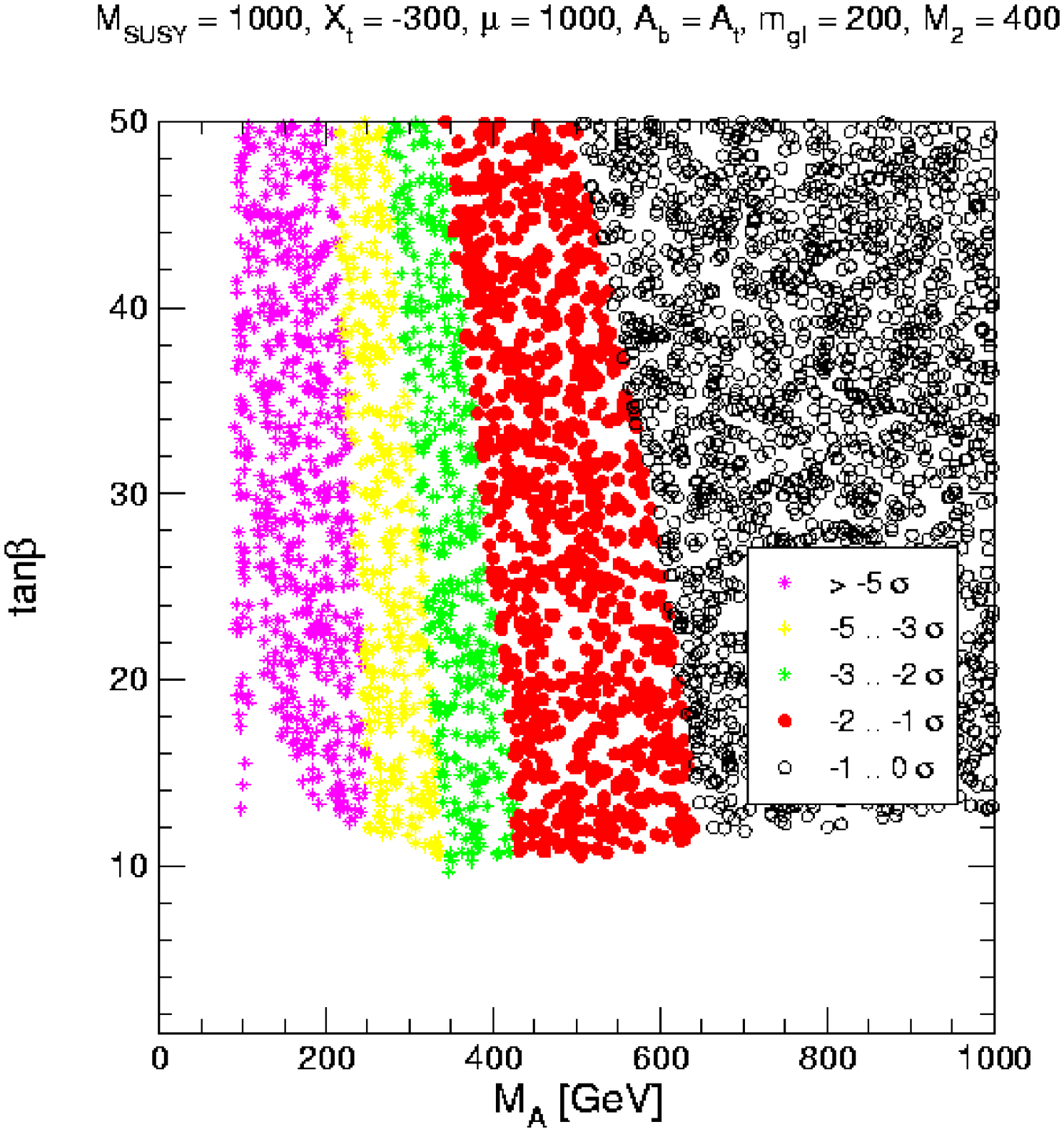,height=7cm}
}
\end{center}
\caption[.]{\label{fig:mssmhgaga}\it
The ratio $R_\ga$ is shown in the $\MA, \tb$ plane for the three benchmark
scenarios described in the text. The observable deviation from the SM
is indicated for a 8\% experimental precision in $R_\ga$.
}
\end{figure}

In Figs.~\ref{fig:mssmhbb}-\ref{fig:mssmhgaga}, the unconstrained MSSM is
analyzed in three benchmark scenarios~\cite{bench}, originally proposed in
connection with Higgs searches at LEP. The `\mhmax' scenario maximizes the
$\mh$ value for a given $\MA, \tb$ combination at fixed $\msusy$ and
$m_t$. The `no-mixing' scenario has the same parameters as the \mhmax\
scenario, but no scalar top mixing. Contrary to \cite{bench}, we have set
$\msusy = 1500$~GeV and $m_{gl} = 1000$~GeV so as to increase the $m_h$ 
values.
In the last scenario, the value of $\mu$ is chosen to be large: $\mu =
1$~TeV. Contrary to \cite{bench}, we have again set $\msusy = 1000$~GeV 
so as to
increase the $m_h$ values. The limits of LEP Higgs searches 
have been applied in these figures,
using an updated version of the results as presented in~\cite{barete}. 
However, the `no-mixing' and 
`large~$\mu$' scenarios have $m_h \lsim 120$~GeV for the shown
parameter space, whereas the \mhmax\ scenario results in 
$m_h \gsim 125$~GeV for large parts of the parameter space, which is
at the limit of the reach of CLICHE.
An analysis of similar scenarios for a TeV-class $e^+ e^-$ linear collider 
can be found in~\cite{lcanalysis}.

In Fig.~\ref{fig:mssmhbb}, $R_b$ is shown in the $\MA, \tb$ plane. As
expected, due to the enhanced $hb\bar b$ coupling, $R_b$ is enhanced in
most part of the parameter space in all three benchmark scenarios.
Suppression only occurs for very small values of $\MA$: $\MA \lsim
130$~GeV or small $\tb$, $\tb \lsim 5$. Deviations from the Standard Model
at the 3-5$\sigma$ level can be observed up to $\MA \lsim 300-500$~GeV,
depending on the scenario. We note deviations from the Standard Model at
the 1$\sigma$ level in nearly the whole plane in the \mhmax\ scenario and
up to $\MA \lsim 500-600$~GeV in the other scenarios. These sensitivities
are in the same ballpark as for a linear  $e^+e^-$
collider~\cite{lcanalysis}. 
Thus, a $\ga\ga$~collider could
offer a complementary method of distinguishing the MSSM from the SM in
the Higgs sector.

Figs.~\ref{fig:mssmhww} and \ref{fig:mssmhgaga} show $R_W$ and $R_\ga$,
respectively, for the three benchmark scenarios. As one would expect from
the enhancement in $R_b$, these two channels are usually suppressed.  
There are cases in which $R_W$ and $R_\ga$ are strongly enhanced 
(by a factor three or four), though these cases are not typical.
Extremely strong suppressions can occur in all scenarios for $\MA \lsim
300$~GeV, rendering these channels more difficult to observe. Deviations
from the Standard Model at the 1$\sigma$ level or better can be found up
to $\MA \lsim 600-1000$~GeV. Thus, these channels could also offer interesting
opportunities to find deviations from the Standard Model over a wide range
of MSSM parameter space.

Contrary to the unconstrained MSSM, where analyses have to be restricted
to certain benchmark scenarios, the full parameter space can be explored
in the CMSSM, where the soft supersymetry-breaking scalar masses $m_0$ and
fermion masses $m_{1/2}$ are assumed to be universal at some input
unification scale. Here we restrict ourselves to $A_0 = 0$ and positive
$\mu$. An exhaustive study can be found in~\cite{ehow2}, and a similar
study for the Tevatron and the LHC is given in~\cite{ehow}.

\begin{figure}[htbp]
\begin{center}\resizebox{\textwidth}{!}{
{\epsfig{file=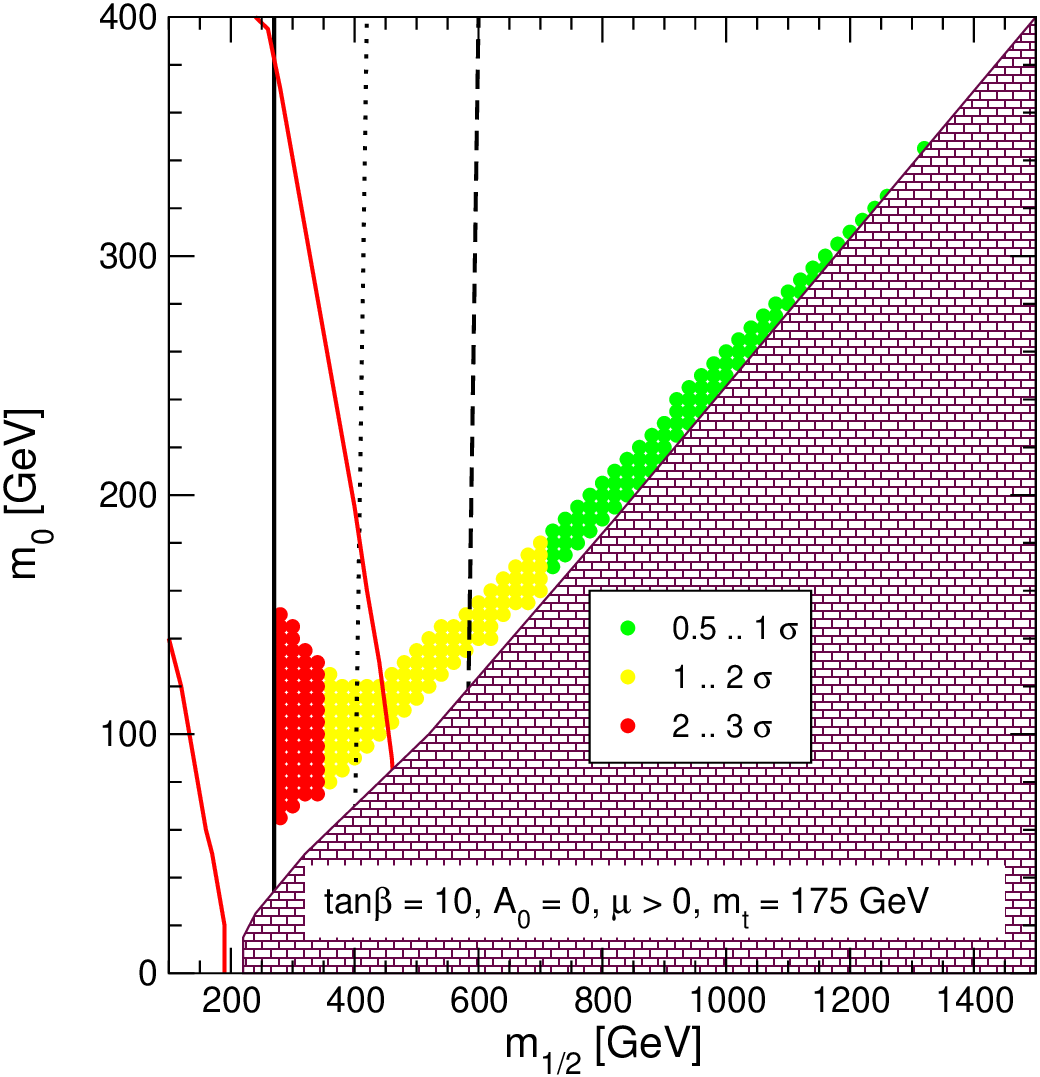,height=8cm}}
{\epsfig{file=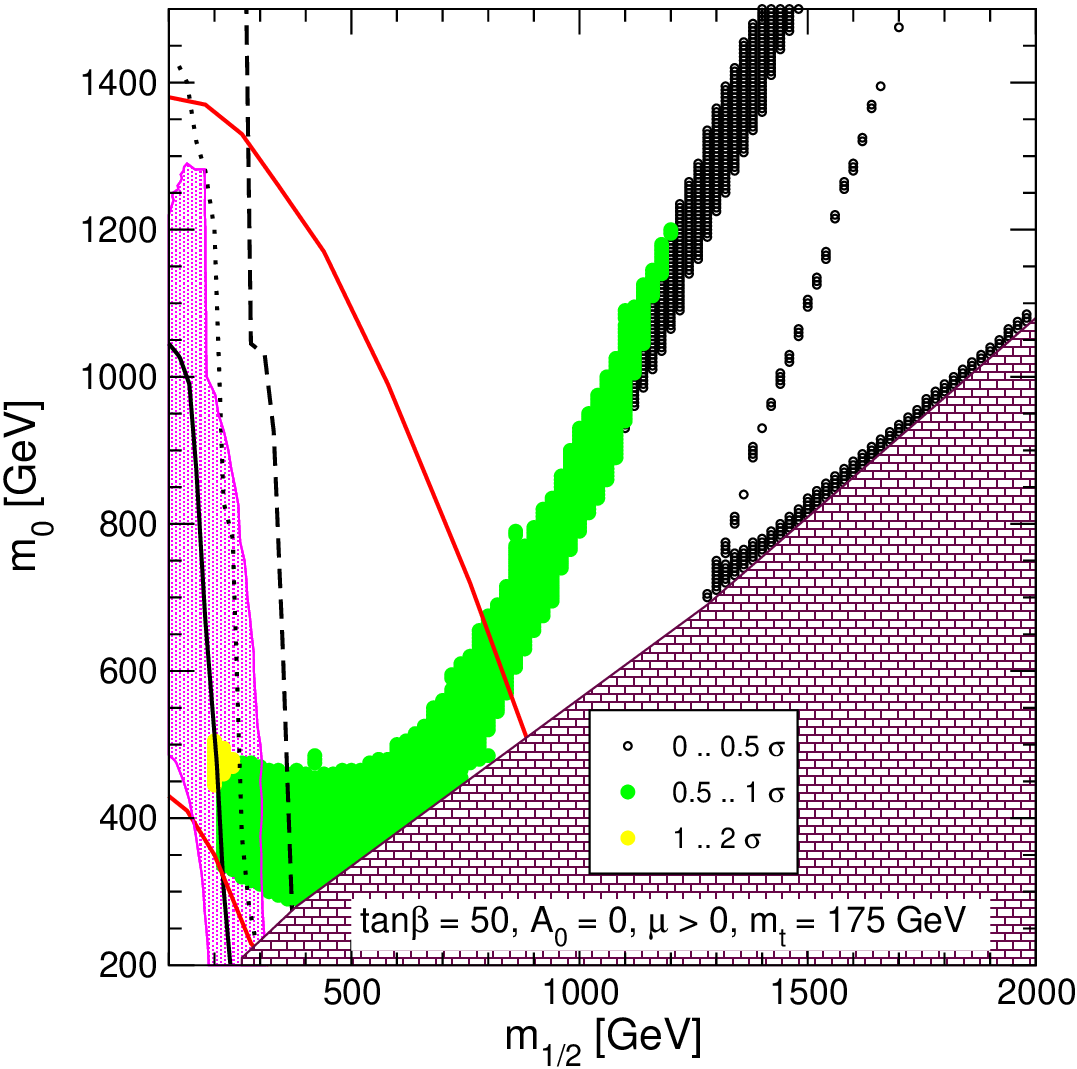,height=8cm}}}
\end{center}
\caption[.]{\label{fig:cmssmhbb}\it
the ratio $R_b$ is shown in the $m_0, m_{1/2}$ plane for
$\tan \beta = 10, 50$ and $\mu > 0$, $A_0 = 0$.
The diagonal (red) solid lines are the $\pm 2 \sigma$ contours for
$g_\mu - 2$.
The near-vertical solid, dotted and dashed (black) lines are the
$\mh = 113, 115, 117$~GeV contours.
The light shaded (pink) regions are excluded
by $b \rightarrow s \gamma$. The (brown) bricked
regions are excluded since in these regions the lightest sparticle
is the charged $\tilde\tau_1$. The observable deviation from the Standard
Model 
is indicated for a 2\% experimental precision of $R_b$.}
\end{figure}

\begin{figure}[ht]
\begin{center}\resizebox{\textwidth}{!}{
{\epsfig{file=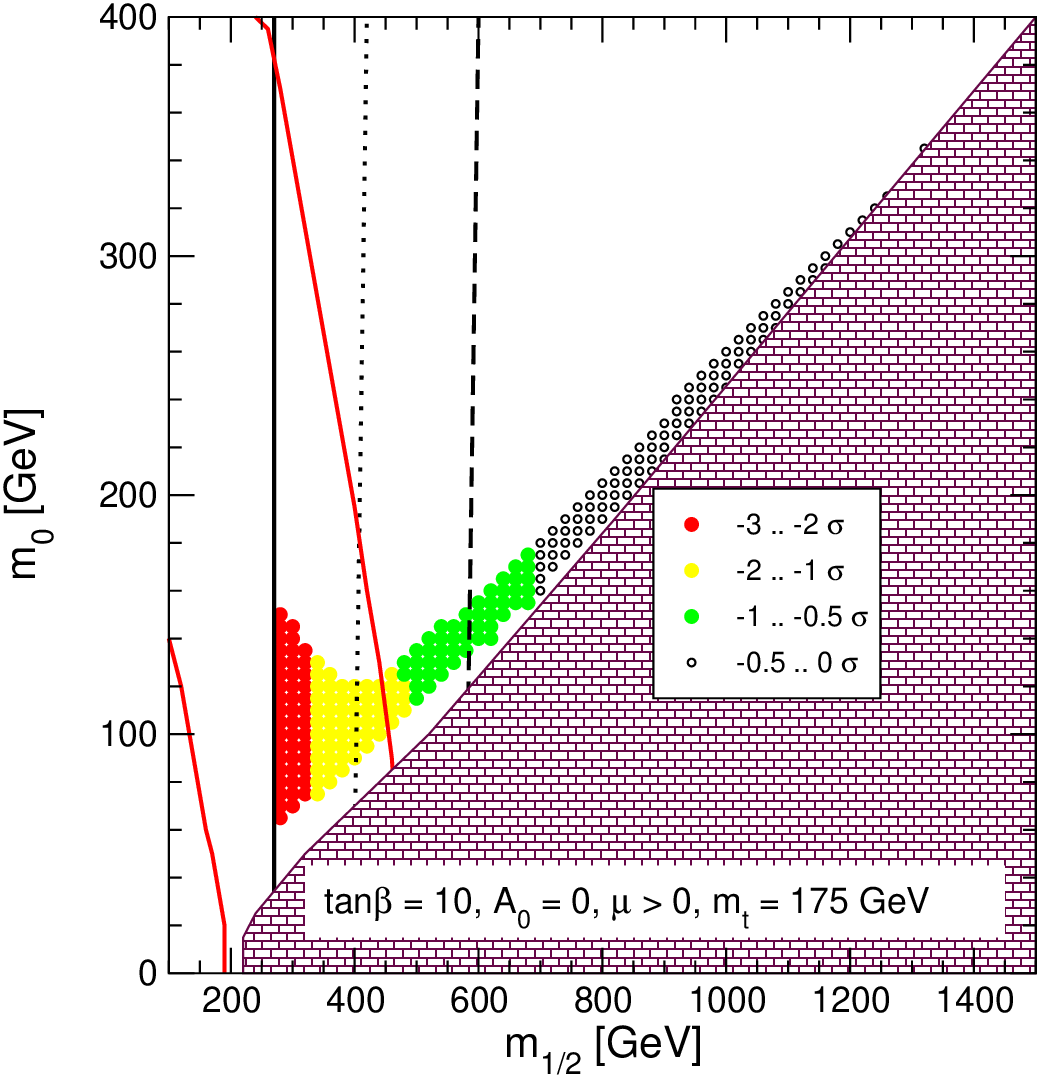,height=8cm}}
{\epsfig{file=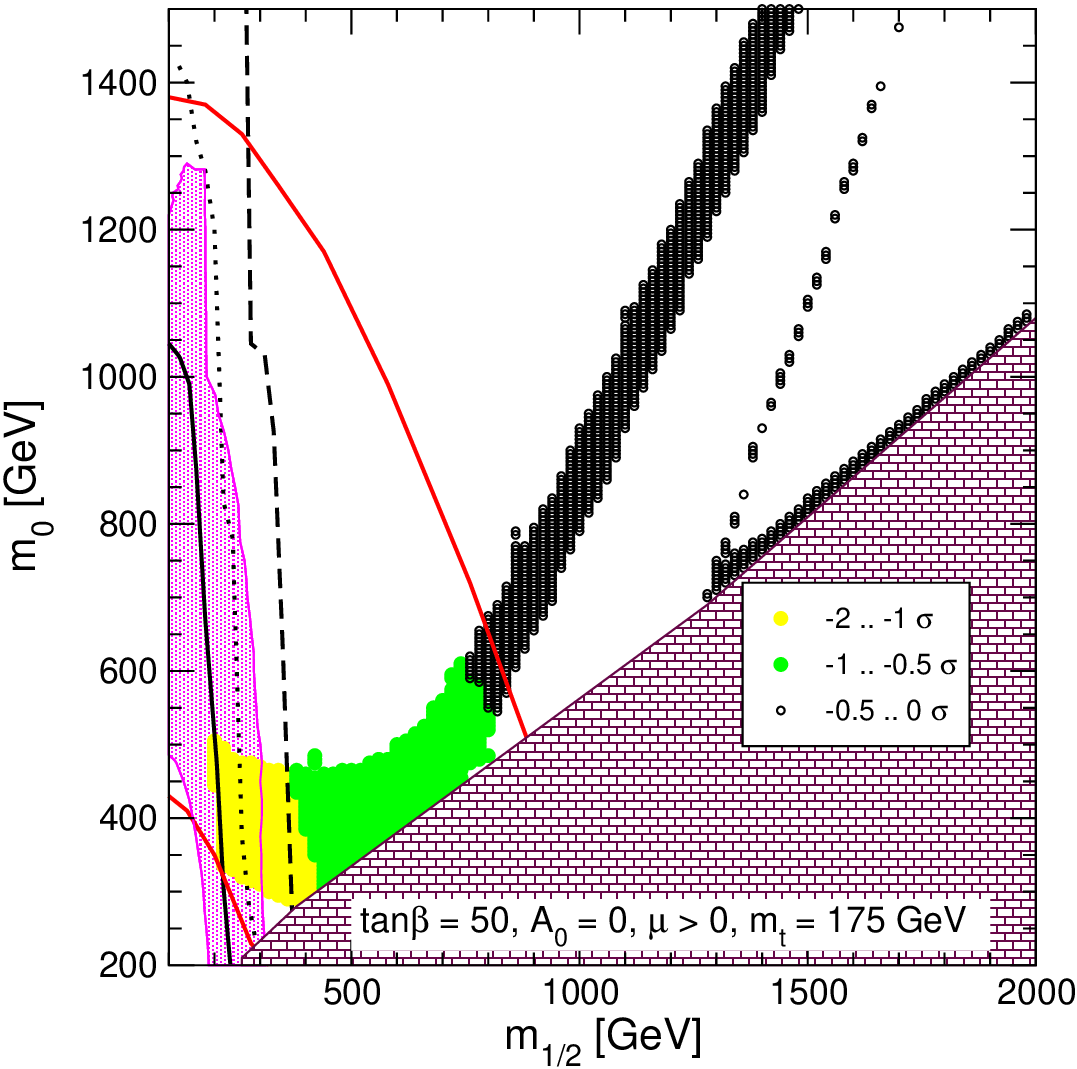,height=8cm}}}
\end{center}
\caption[.]{\label{fig:cmssmhww}\it
The ratio $R_W$ is shown in the $m_0, m_{1/2}$ plane for
$\tan \beta = 10, 50$ and $\mu > 0$, $A_0 = 0$.
The diagonal (red) solid lines are the $\pm 2 \sigma$ contours for
$g_\mu - 2$.
The near-vertical solid, dotted and dashed (black) lines are the
$\mh = 113, 115, 117$~GeV contours.
The light shaded (pink) regions are excluded
by $b \rightarrow s \gamma$. The (brown) bricked
regions are excluded since in these regions the lightest sparticle
is the charged $\tilde\tau_1$. The observable deviation from the Standard
Model
is indicated for a 5\% experimental precision of $R_W$.}
\end{figure}

\begin{figure}[ht]
\begin{center}\resizebox{\textwidth}{!}{
{\epsfig{file=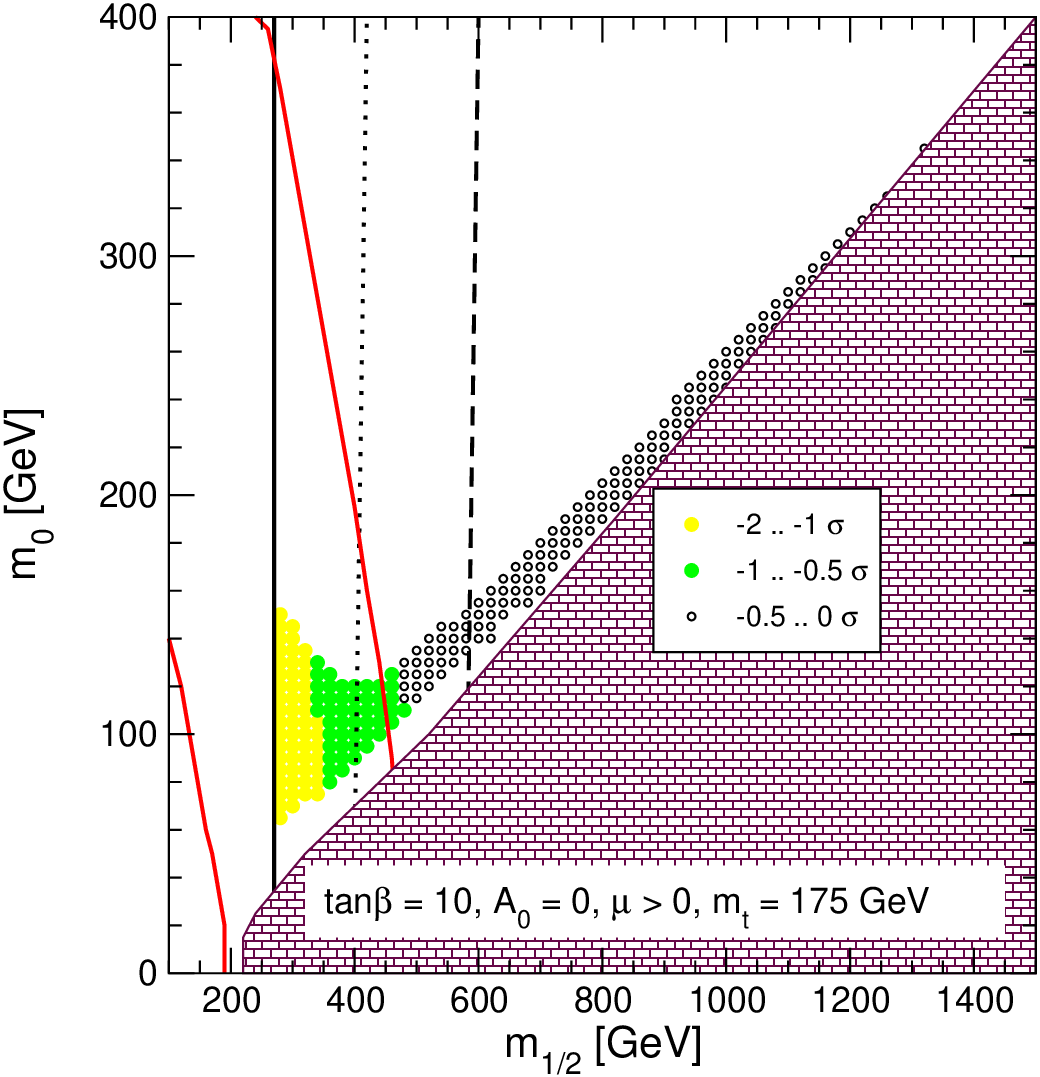,height=8cm}}
{\epsfig{file=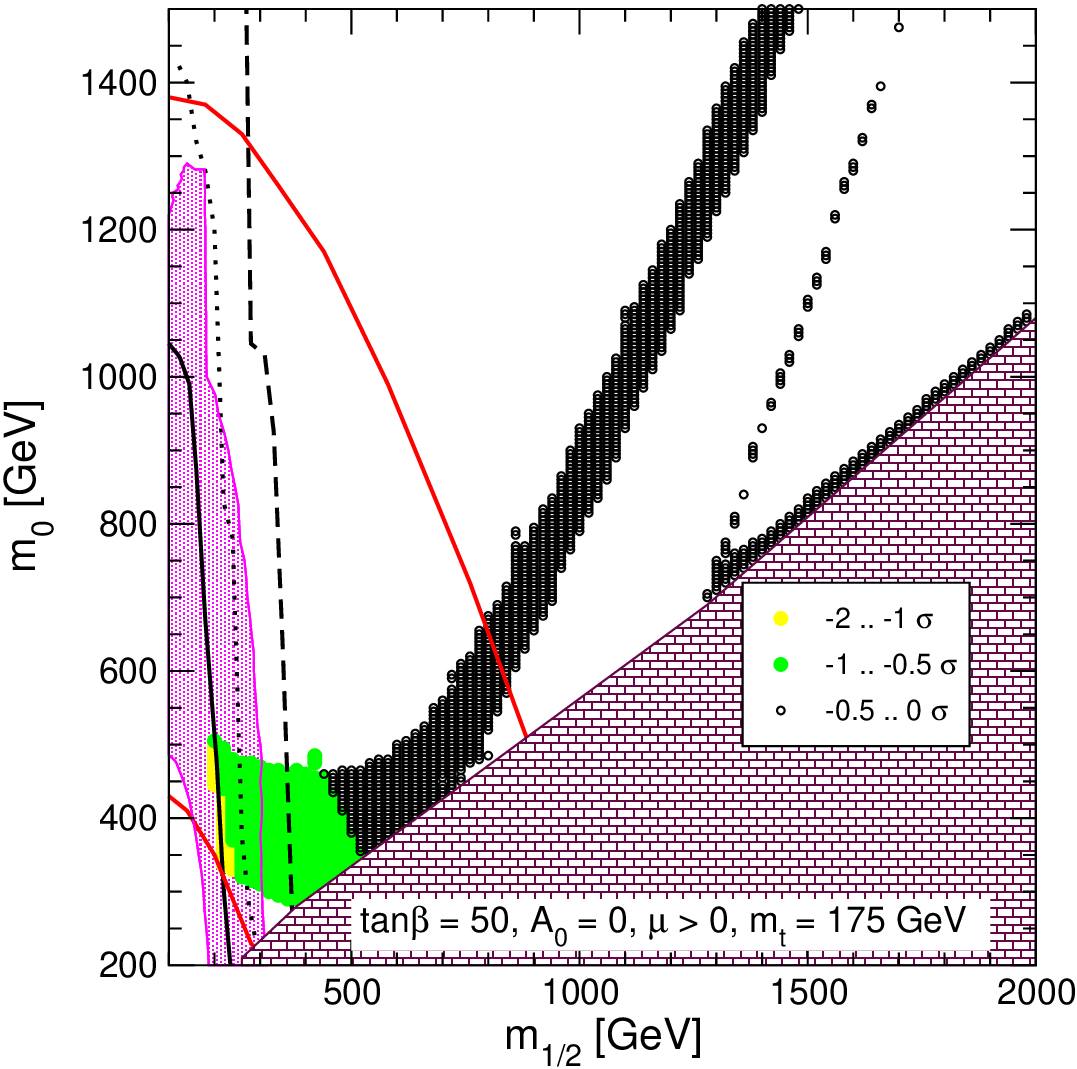,height=8cm}}}
\end{center}
\caption[.]{\label{fig:cmssmhgaga}\it
The ratio $R_\ga$ is shown in the $m_0-m_{1/2}$ plane for
$\tan \beta = 10, 50$ and $\mu > 0$, $A_0 = 0$.
The diagonal (red) solid lines are the $\pm 2 \sigma$ contours for 
$g_\mu - 2$.
The near-vertical solid, dotted and dashed (black) lines are the
$\mh = 113, 115, 117$~GeV contours.
The light shaded (pink) regions are excluded
by $b \rightarrow s \gamma$. The (brown) bricked
regions are excluded since in these regions the lightest sparticle
is the charged $\tilde\tau_1$. The observable deviation from the Standard
Model
is indicated for a 8\% experimental precision of $R_\ga$.}
\end{figure}

The ratios $R_{b,W,\ga}$ are shown in
Figs.~\ref{fig:cmssmhbb}-\ref{fig:cmssmhgaga} in the $m_{1/2}, m_0$ plane
for $\tb = 10, 50$, $\mu > 0$ and $A_0 = 0$. Only the parameter space that
gives acceptable values for a CMSSM explanation of cold dark matter, $0.1
\le \Omega_{CDM} h^2 \le 0.3$, is analyzed: see~\cite{cdm} for details.
Correspondingly, the regions with a $\tilde{\tau}$ LSP are marked as
excluded. In addition, the regions disfavored by measurements of 
${\cal B}r(b \to s \ga)$~\cite{bsg} are indicated, as are the regions 
preferred by the recent $g_\mu - 2$ measurement~\cite{gm2}.

As in the unconstrained MSSM, $R_b$ is enhanced in the CMSSM, whereas 
$R_W$ and $R_\ga$
are suppressed. Larger deviations are observed for lower values of
$m_{1/2}$ and $m_0$, which are also preferred by the $g_\mu-2$
measurement. For $\tb = 10$, up to 3$\sigma$ could be observable, whereas
for $\tb = 50$ the maximal deviation of the MSSM from the SM could be
2$\sigma$. Similar values would be obtainable at a linear $e^+e^-$ 
collider~\cite{ehow2}.
Therefore, since it uses a different production process, a
$\ga\ga$~collider could
provide additional complementary information to the results obtainable at
a linear  $e^+e^-$ collider.

Generally, it is expected that supersymmetry, if it exists, will be
discovered at the LHC via the  production and observation of 
sparticles. However one can construct also 
so-called  Standard-Model-like scenarios where 
only one light Higgs boson will be within the reach 
of the LHC and future linear $e^+e^-$ colliders, and its measured
couplings to quarks,
leptons and
gauge bosons will be in agreement with their SM expectation within 
experimental errors.

Such scenarios can be constructed in MSSM and in more general
two-Higgs-Doublet Models (2HDM), as demonstrated in~\cite{krawczyk}.  In
the latter study, the authors took the CP-conserving 2HDM in its model II
implementation, where one doublet of fundamental scalar fields couples to
the $u$ quarks and the other to the $d$ quarks and charged leptons. The
experimental accuracies with which couplings are expected be measured at a
500~GeV $e^+e^-$ linear collider are taken into account in the definition
of a parameter space in the 2HDM where it would be indistinguishable from
the Standard Model.  Hence, the Higgs mimics all the Standard Model
properties one can expect to be measured using LHC and linear collider
data. 

\begin{figure}[htbp]
\begin{center}
\epsfig{file=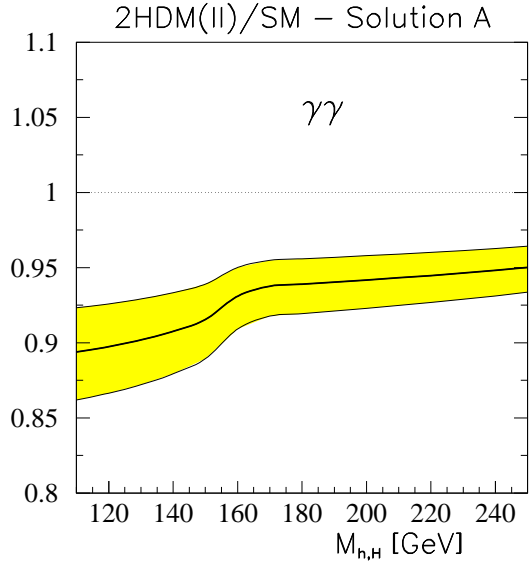,height=8cm}
\end{center}
\caption[.]{\label{fig:maria}\em
Ratios of the Higgs boson $\gamma\gamma$-decay
width in the 2HDM and the Standard Model as functions of $m_{h,H}$,
assuming that all basic  couplings are indistinguishable from those in the
Standard Model~\cite{krawczyk}.
}
\end{figure}

Measuring the partial width of the Higgs to photons can distinguish
between such scenarios.  Due to the contribution to the
$H\gamma\gamma$ coupling of all charged particles, including the very
heavy ones, the ratio of the Higgs boson width in the 2HDM to the one in
the Standard Model can differ significantly from unity.  An
example is shown in Fig.~\ref{fig:maria} for a 2HDM solution which
satisfies all the Standard-Model-like criteria.  The possible deviation
from the Standard Model for $m_H \sim 120$~GeV is considerably larger than
the error in the $\gamma \gamma \rightarrow H \rightarrow {\bar b} b$
signal expected from CLICHE.  This stresses the importance of accurate
measurements of the two photon width of the Higgs, and the
correspondingly unique role of
a precision photon collider in disentangling physics beyond the Standard
Model.

\subsection{QCD Physics in $\gamma\gamma$ Collisions}

QCD aspects of $\gamma\gamma$ physics have been studied at $e^+e^-$
colliders over the last 20 years. At LEP, $\gamma\gamma$ collisions with
$\sqrt{s}(\gamma\gamma)$ up to 140~GeV have been studied. Up to now, the
photons have been produced via bremsstrahlung~\cite{wws} from the electron
and positron beams, leading to soft energy spectra with only limited
statistics at high $\sqrt{s}(\gamma\gamma)$, whereas CLICHE will produce
$\gamma\gamma$ collisions in the high-energy part of the spectrum. A
plethora of QCD physics topics in two-photon interactions can be addressed
with a $\gamma\gamma$ collider, as recently discussed in~\cite{teslatdr}.
Furthermore, good knowledge and understanding of two-photon processes will
be essential for controlling physics background contributions to other
processes and machine backgrounds at TeV and multi-TeV linear $e^+e^-$
colliders.

A key issue is the total $\gamma\gamma$ cross section, which is not yet
understood from first principles. Present data show a rise in
$\gamma\gamma$ collisions that may be faster than that in $pp$, but the
experimental errors are still large. At a $\gamma \gamma$ collider such as
CLICHE, the photon beam energy can be tuned with a spread of less than
10\%, so that measurements of $\sigma_{tot} (\gamma \gamma)$ can be made
at a number of `fixed' energy values. The absolute precision with which
these cross-sections can be measured ranges from 5\% to 10\%, according to
studies made for  the $\gamma \gamma$ option of  TESLA~\cite{teslatdr}.

Quantum fluctuations of the photon into quarks or bound states lead to the
so-called hadronic structure of the photon. The absolute magnitude of the
photon hadronic structure function is asymptotically determined by the
strong coupling constant~\cite{gg_witten}. The classical way to study the
structure of the photon is via deep inelastic electron-photon scattering, 
i.e., two-photon interactions with one quasi-real (virtuality $Q^2
\sim 0$) and one virtual ($Q^2 >$ few~GeV$^2$) photon, which can be
achieved by switching off one the laser beams. Making the reasonable
assumption that the scattered electron can be detected down to 25~mrad,
measurements can be made in the region $5.6 \cdot 10^{-4} < x < 0.56$,
where $x$ is the fraction of the photon momentum carried by a constituent
parton, and $10 < Q^2 < 8\cdot 10^{3}$~GeV$^2$.

Although e$\gamma$ scattering allows one to measure the quark
distributions inside the photon, it constrains only weakly the gluon
distribution, via the QCD evolution of the structure functions. Direct
information on the gluon in the photon can, however, be obtained from
measurements of jets~\cite{wengler}, open charm~\cite{jankowski} and
$J/\psi$~\cite{indumathi} production in $\gamma\gamma$ interactions at an
$e\gamma$ and $\gamma\gamma$ collider. Values of $x$ down to a few $\times
10^{-3}$ can be reached with charm and di-jet
measurements~\cite{wengler,jankowski}, a region where predicted gluon
distributions typically differ by a factor of two or more.

We also recall the deviation~\cite{L3bbar,OPALbbar} from the NLO QCD 
predictions of the
$b\overline{b}$ cross section in $\gamma\gamma$ collisions measured at
LEP, which was mentioned earlier. It is unlikely that this matter will be
settled by further analysis of the LEP data, and CLICHE could revisit the
study of the $b\overline{b}$ cross section.  It will allow accurate
measurements as functions of $W_{\gamma\gamma}$ and other kinematical
variables to identify the origin of the putative anomaly.

A linear collider also provides circularly-polarized photon beams, which
offer a unique opportunity to study the polarized parton distributions of
the photon, for which no experimental data are available so far.  
Information on the spin structure of the photon can be obtained from
inclusive polarized deep-inelastic e$\gamma$ measurements and from jet and
charm measurements~\cite{stratmann,g1_kwiecinski} in polarized $\gamma
\gamma $ scattering.  Measurements of $g_1$, particularly at low $x$, are
very important for studies of the high-energy QCD limit, where signs of
the BFKL regime~\cite{bfkl} may appear.

Other dedicated measurements have been proposed for detecting and studying
the large $\ln1/x$ logarithm resummation effects in QCD. One example is
vector meson production, e.g., $\gamma \gamma $ $\rightarrow J/\psi
J/\psi$ or (at large $t$)  $\gamma \gamma \rightarrow \rho\rho$, where the
hard scale in the process is given by the $J/\psi$ mass or the momentum
transfer $t$. The $J/\psi$ can be detected via its decay into leptons, and
separated from the background via its peak in the $\ell^+ \ell^-$
invariant mass.  Other processes that are strongly sensitive to BFKL
effects include e$\gamma$ scattering with associated jet
production~\cite{contreras}, and $e^+e^- \rightarrow e^+e^-\gamma X$ and
$\gamma\gamma \rightarrow \gamma X$~\cite{evanson}.

\subsection{$e^- \gamma$ Physics}

\begin{figure}[t]
\begin{center}
\mbox{\epsfig{file=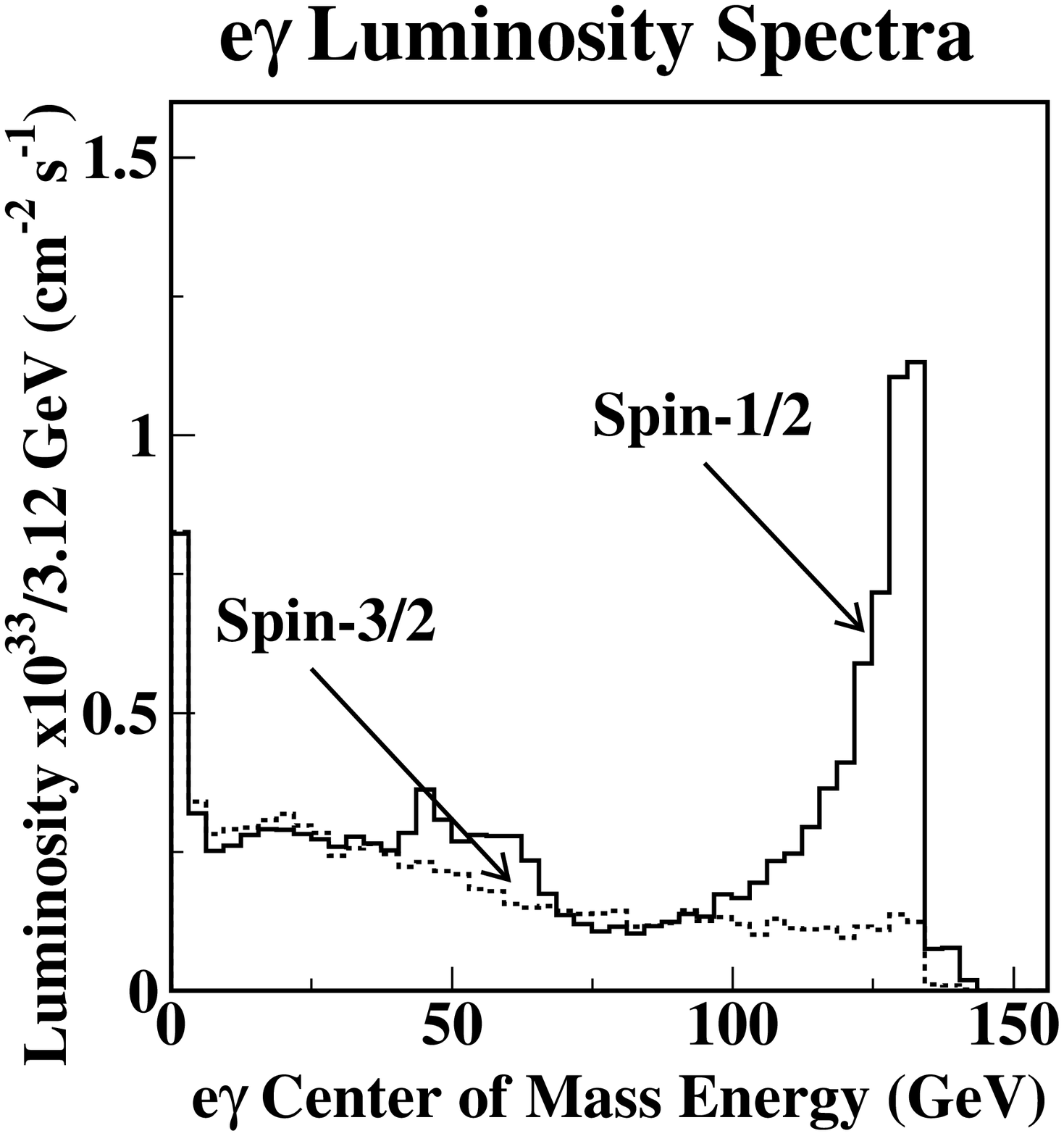,height=8cm}}
\mbox{\epsfig{file=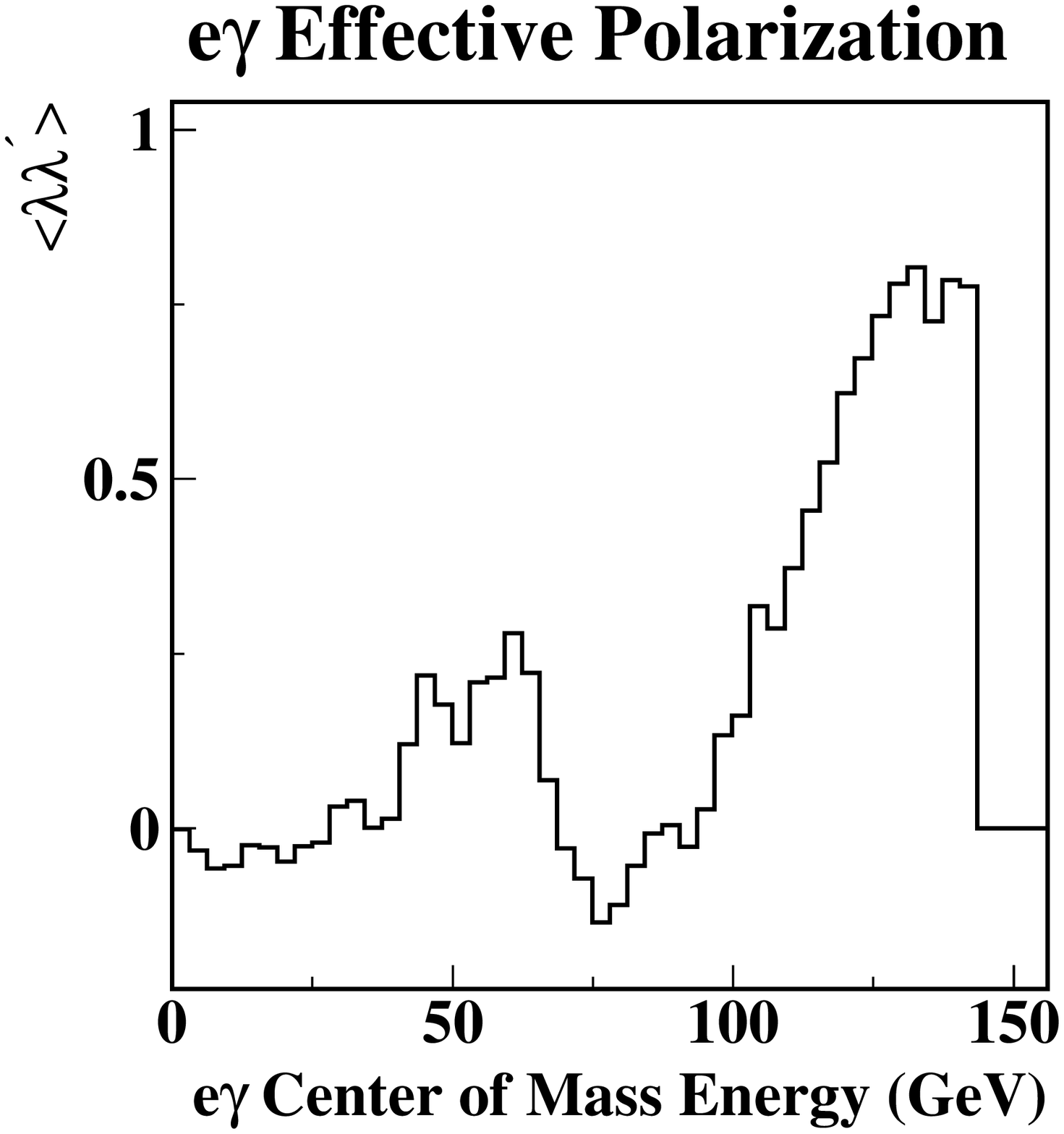,height=8cm}}
\end{center}
\caption[.]{\label{fig:eg}\it
Luminosity spectra  and polarization for different spin states as functions of
$E_{CM}$($e^-\gamma$), assuming the CLIC~1 parameters for 75~GeV electrons
obtained with 
{\tt DIMAD}~\cite{dimad} and {\tt CAIN}~\cite{cain2} for 
${{\cal L}_{ee}=4.8 \times10^{34} \rm cm^{-2}s^{-1}}$.

}
\end{figure}

\begin{figure}[htbp]
\begin{center}  
\mbox{\epsfig{file=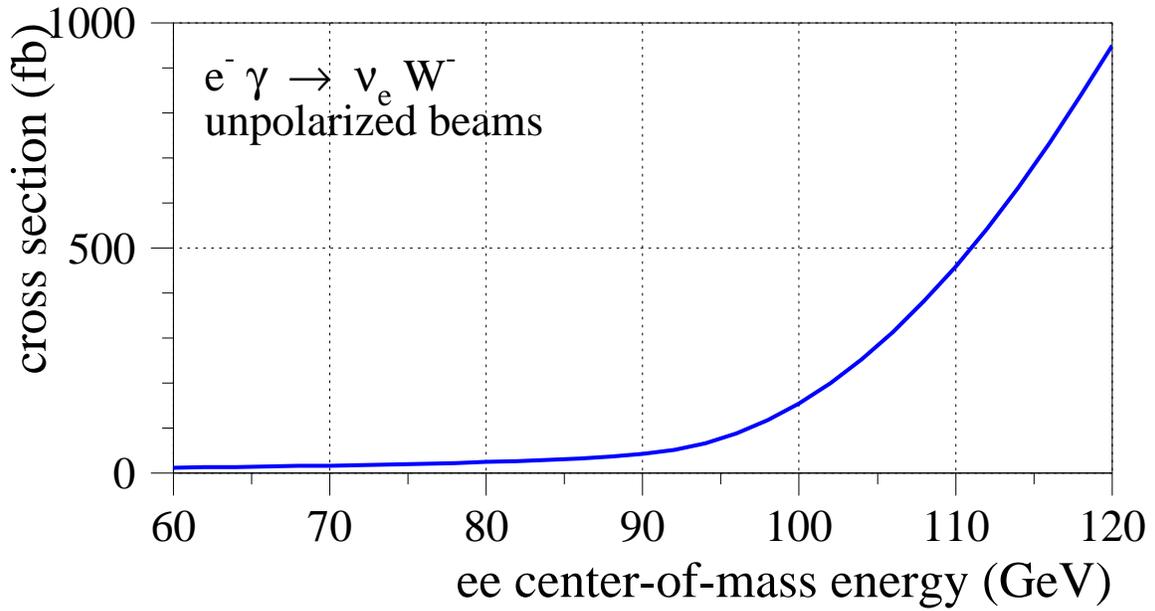,height=8cm}}
\end{center}
\caption[.]{\label{fig:Wthreshold}\it
The rise of the cross section for $e^- \gamma \rightarrow \nu W^-$ as a
function of $E_{CM}(e^-e^-)$.
}
\end{figure}

\begin{figure}[htbp]
\begin{center}  
\mbox{\epsfig{file=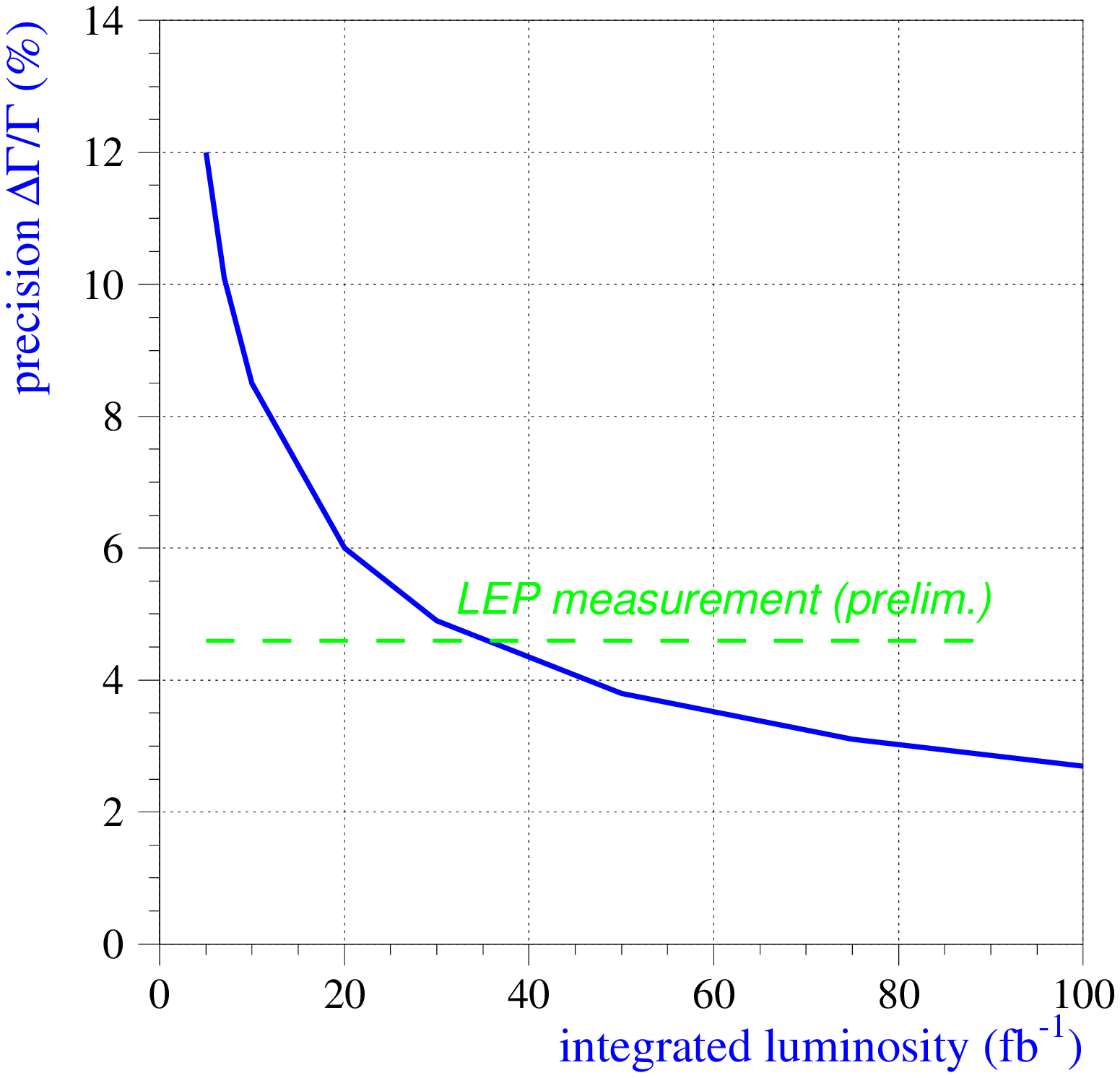,height=8cm}}
\end{center}
\caption[.]{\label{fig:GammaW}\it
The sensitivity with which $\Gamma_W$ might be measured at in 
$e^-\gamma$ collisions, as a function of the integrated luminosity available.
}
\end{figure}

The option of $e^- \gamma$ collisions is available along with $\gamma
\gamma$ and $e^- e^-$ collisions, because the $e^-$ conversion efficiency 
is, by design, less than 100\%. The expected luminosity spectra for 
different spin states provided by the CLICHE design described previously 
are shown in Fig.~\ref{fig:eg}.
Moreover, if one wanted 50\%  higher $e^- \gamma$ luminosity at the peak, 
one could switch off one of the laser backscattering systems at cost 
of a factor of two in the total  $e^- \gamma$ luminosity.
Among the reactions of
potential interest, we mention $e^- \gamma \rightarrow \nu W^-$. As seen
in Fig.~\ref{fig:Wthreshold}, the cross section for this process rises
rapidly with $E_{CM} (e^- e^-)$ in the range accessible to CLICHE. This
reaction could in principle be used to measure $m_W$ and/or $\Gamma_W$.
The result of one exploratory study is shown in Fig.~\ref{fig:GammaW}. It
shows the accuracy attainable in a measurement of $\Gamma_W$ as a function
of the available integrated $e^- \gamma$ luminosity. We see that a CLICHE
measurement could become competitive for a luminosity of 50~fb$^{-1}$ or
more. 

\section{Outlook}

The technology to build a machine like CLIC is under active development,
but is not available today.  Important progress in establishing the
two-beam acceleration technique, as a novel method to obtain large
gradients, has been made over the last 5 years with the CLIC Test
Facilities~1 and 2. Presently, a new test facility is being prepared to
demonstrate the principle of all ingredients needed to build up the
required drive beam. The final results of this test facility are expected
by the end of 2006, and, if successful, this could initiate the preparation
of a Technical Design Report for a machine based on CLIC technology.  As
mentioned in the Introduction to this paper, the next technological step
towards a multi-TeV collider could be the construction and operation of
one (or, as we suggest, two) full CLIC module(s),
providing acceleration by about 70~GeV.
By the time one would have to define the
physics objectives of such a CLIC~1 stage, it will be clear from the
Tevatron and LHC if the Higgs exists and is within the mass reach of 
CLICHE. 

In this scenario, CLICHE could be contemporary with the operation of a
TeV-class $e^+e-$ linear collider such as TESLA, NLC or JLC. The
complementary information on the Higgs boson provided by CLICHE could be
very valuable and help to distinguish among models. 

All the above linear collider proposals consider a $\gamma\gamma$ collider
as an option that could be added to their baseline programmes.  The
physics programme of a higher-$E_{CM}$ $\gamma \gamma$ collider has been
amply documented~\cite{nlc_report,teslatdr}. Here we just recall that 
such a machine could provide a
unique window on the heavier neutral Higgs bosons $H, A$ expected in the
MSSM and 2HDM~\cite{muhl,david_jack,krawczyk}, 
and would offer bright prospects for unravelling their CP
properties.  The high-energy physics programme for an $e^+e^-$ collider is
by itself so rich that one can expect any photon collider option to start
only several years of the start-up of the facility. The experience on
photon colliders that could be gained earlier at a dedicated facility such
as CLICHE could be exploited at the higher energies attainable at a
TeV-scale linear collider, and eventually also at a multi-TeV collider
such as CLIC~\cite{MultiTeVGG}. 

Clearly the exploratory studies on machine and physics presented here need
to be pursued with more detailed analyses.  Ideas exist on the machine
side that may lead to an increased luminosity  for CLICHE. Also, it
is possible that the number of lasers needed could be reduced by using
recirculating laser pulses in the interaction region.  If the Higgs proves
to be heavier than about 125~GeV, one could also think of upgrading
the input beam energy to CLIC~1. 

We recall that CLICHE is just one of several possible options for doing
physics with CLIC~1, many of which are more conventional and 
deserving serious study. However, we consider CLICHE to be a very
attractive option for a project that could simultaneously validate and
test all components of the CLIC technology for accelerating high-energy
beams and can give important scientific output, covering a unique facet of
the study of the Higgs boson, whose study will be central to physics at
the high-energy physics frontier over the next decade or two. 

\section{Acknowledgement}

This work was performed in part under the auspices of the U.S. Department of 
Energy by the University of California, Lawrence Livermore National
Laboratory under Contract No.W-7405-Eng.48, and the 
Illinois Consortium for Accelerator Research, ICAR.

\end{document}